\def\|{\kern3pt}

\def\LEM #1. #2\par{\medbreak
 \noindent{{\it \bf Lemma\ #1.}\enspace}{\sl#2\par}%
  \ifdim\lastskip<\medskipamount \removelastskip\penalty55\medskip\fi}

\def\PROP #1. #2\par{\medbreak
 \noindent{Proposition.\kern8pt #1\enspace}{\sl#2\par}%
 \ifdim\lastskip<\medskipamount \removelastskip\penalty55\medskip\fi}

\def\consequence #1. #2\par{\medbreak
  \noindent{{\bf Corollary #1}.\enspace}{\sl#2\par}%
  \ifdim\lastskip<\medskipamount \removelastskip\penalty55\medskip\fi}

\def\df #1. #2\par{\medbreak
  \noindent{{\tt {\bf Definition#1.}}\enspace}{\sl#2\par}%
  \ifdim\lastskip<\medskipamount \removelastskip\penalty55\medskip\fi}

\def\note{{\sl\bf \medbreak\noindent Remark.\ \ \ }}
\def\one{{\bf1}}

\def\dsl{\raise.15ex\hbox{/}\kern-.57em\partial}  
\def\xample{\medbreak\noindent {\it Example }}
\def\proof{\medbreak\noindent {\it Proof.\ \ \ }}
\def\ss#1#2{\mathrel{\mathop{#2}\limits^{#1}}}

\def\dal{{\partial_0}^2-{\partial_1}^2-{\partial_2}^2-{\partial_3}^2}
\def\fin{\hbox{$\ {}_\bullet$}}
\def\4{$4\!\times\!4$}
\def\gc{\Gamma_{\cal C}}
\def\S{{\cal S}}
\def\V{{\cal V}}
\def\exp{{\rm exp}}
\def\com{{\rm com}}
\def\Spin{{\rm Spin}}

\def\diag{{\rm diag}}
\def\tr{{\rm tr}}
\def\U{{\rm U}}
\def\Sp{{\rm Sp}}
\def\SU{{\rm SU}}
\def\SO{{\rm SO}}
\def\su{{\rm su}}
\def\sp{{\rm sp}}
\def\Ad{{\rm Ad\,}}
\def\com{{\rm com}}
\def\u{{\rm u}}
\def\gnumbers{$\gamma$-numbers}

\def\g{$\gamma$-}

\documentstyle[numreferences]{aacalate}
\begin{opening}
\title
{DIRAC \g EQUATION, CLASSICAL GAUGE FIELDS
AND CLIFFORD ALGEBRA
\footnote[1]{\vskip-.25cm\hskip-.3cm {\it
Advances in Applied Clifford Algebras} {\bf 8} {\rm No. 1, 181-224
(1998)}}}

\author{N. G. Marchuk}
\institute{Steklov Mathematical Institute, \\
Gubkina st. 8,\\
  Moscow GSP-1, 117966, Russia \\
email: nikolai@marchuk.mian.su}
\date{(Received: April 3, 1998, \quad Accepted: April 7, 1998)}
\end{opening}

\runningtitle{\underline{Advances in Applied Clifford
Algebras 8, No. 1 (1998)}}
\runningauthor{\underline{
Dirac \g Equation, Classical Gauge Fields
and Clifford Algebra\qquad\hfill N. G.Marchuk}}

\begin{document}

\hsize=4.7in
\vsize=6.5in
\voffset=.3in

\font\eightrm=cmr8
\font\eightbf=cmbx8
\font\eightit=cmti8
\font\eightsl=cmmi8

\font\twelverm=cmr12
\font\twelvebf=cmbx12
\font\twelveit=cmti12
\font\twelvesl=cmmi12

\def\medtype{\let\rm=\eightrm \let\bf=\eightbf
\let\it=\eightit \let\sl=\eightsl \let\mus=\eightmus
\baselineskip=10.5pt
\rm}

\def\bigtype{\let\rm=\twelverm \let\bf=\twelvebf
\let\it=\twelveit \let\sl=\twelvesl \let\mus=\twelvemus
\baselineskip=10.5pt
\rm}

\def\qed{\hfill\unskip\nobreak\quad\qedrule\medbreak}

\def\square{\leavevmode\thinspace\hbox{\vrule\vtop{\vbox{\hrule\kern0pt
\hbox{\vphantom{\tt/}\thinspace{\tt$\, \,$}\thinspace}}
\kern0pt\hrule}\vrule}\thinspace}

\def\uno{{\bf 1}}
\def\P{I\!\!P}
\def\R{I\!\!R}
\def\C{I\!\!\!\!C}
\def\N{I\!\!N}
\def\M{I\!\!M}
\def\La{I\!\!L}
\def\F{I\!\!F}
\def\Z{Z\!\!\!Z}
\def\X{\bf X\!\!\!\!\!X}
\def\HH{I\!\!H}
\def\W{W\!\!\!\!\!\!W}

\font\sqi=cmssq8
\def\DU{\rm 1\kern-2.95pt\rm I}
\def\DR{\rm I\kern-2.45pt\rm R}
\def\DC{\kern2pt {\hbox{\sqi I}}\kern-4.2pt\rm C}
\def\MDC{{\cal M}\kern2pt {\hbox{\sqi I}}\kern-4.2pt\rm C}
\def\DN{\rm I\kern-1.45pt\rm N}
\def\DZ{\hbox{$\rm Z$}
        \hbox{\kern -4pt \rm Z}}
\def\DZP{\hbox{$\sty Z$}
        \hbox{\kern -4pt $\sty Z$}}

\def\BR{{\rm I\kern-2pt\rm R}}
\def\BD{{\rm I\kern-2pt\rm D}}
\def\BH{{\rm I\kern-2pt\rm H}}

\def\small{\let\rm=\eightrm \let\bf=\eightbf
\let\it=\eightit \let\sl=\eightsl \let\mus=\eightmus
\baselineskip=10.5pt
\rm}

\def\tit #1\par{\medbreak
\noindent{{\bf#1}}\par\medbreak\no}

\def\ti2 #1. #2\par{\medbreak
\noindent{\bf#1. \enspace}{\sl#2}\par\medbreak\no}

\def\2ti #1.#2. #3\par{\medbreak
\noindent{\bf#1.#2.} \enspace {\sl#3}\par\medbreak\no}

\def\TI #1. #2\par{\medbreak
\noindent{\bf#1. \enspace}{\rm#2}\par\medbreak\no}

\def\ab \par{\medbreak
\centerline{\bf Abstract}\par\medbreak}
\def\subsec #1. #2. #3\quad{\medbreak
\noindent{\bf#1.\enspace}{\bf#2.\enspace}{\bf#3.\enspace}\quad}

\def\dj{d\kern-.35em\raise1.25ex\vbox{\hrule width .3em height .03em}}
\def\Dj{D\kern-.70em\raise0.75ex\vbox{\hrule width .3em height .03em}
\kern.03em}

\def\und{\underbar}
\def\no{\noindent}
\def\unme{{1\over 2}}
\def\ii {\'{\i}}

\def\qedrule{\hbox{\vrule height1.4ex depth0pt width1ex}}
\def\qed{\unskip\nobreak\quad\qedrule\medbreak}
\def\E#1#2#3{E\,_{#1}\,^{#2}\,_{#3}}


\def\hv#1{\hskip-0.5em_{_{_{_{_{#1}}}}}}

\def\hv#1{\hskip-2.5em_{_{_{_{_{#1}}}}}}

\baselineskip11.8pt
\def\C{{\cal C}}
\def\M{{\cal M}}

\begin{abstract}
 An equation, we call Dirac \g equation,
is introduced with the help of the mathematical tools connected
with the Clifford
algebra. This equation can be considered as a generalization of
the Dirac
equation for the electron.  Some features of Dirac \g equation are
investigated (plane waves, currents, canonical forms).
Furthermore,  on the basis
of local gauge invariance regarding unitary group, a system of
equations
is  introduced consisting of Dirac \g equation and the Yang-Mills
or Maxwell equations. This system of equations describes a Dirac's
field  interacting  with
the Yang-Mills or Maxwell gauge field.  Characteristics of this
system of equations are studied for various gauge groups and the
liaison between the new  and the standard constructions of
classical gauge fields
is discussed.
\end{abstract}
\newpage

\tableofcontents

\section{Introduction.}

 Soon after the finding of Dirac's famous equation
for the electron  in 1928 [10], several papers appeared  H. Weil [1] and
V. A. Fock [2], among others, in which  the
electromagnetic  field  was  described  as a gauge field of the
Dirac equation required from
demanding of local gauge invariance with respect to a phase
transformation
(Abelian U(1) gauge group) of the Dirac's lagrangian.  Further development
of this approach allowing for the non-Abelian gauge fields after
the  work  of  Yang  and Mills (1954) [3] considering the group of
``isotopic'' transformations (gauge group SU(2)).  Their work was  soon
generalized to a wider class of Lie groups. Non-Abelian gauge fields
began to be named Yang-Mills fields, whereas the equations describing
them  were named Yang-Mills equations.  In modern physics such fields
are used in models of electroweak and strong interactions.
     In the present article an equation, we call Dirac \g equation,
is introduced with the help of the mathematical tools connected  with
the Clifford
algebra. This equation can be considered as a generalization of Dirac
equation for the electron.  Certain features of Dirac \g equation are
investigated (plane waves, currents, canonical forms). Further on the basis
of local gauge invariance regarding unitary group a system of  equations
is  introduced consisting of Dirac \g equation and equations
of Yang-Mills or Maxwell type. That system of equations describes Dirac's
field  interacting  with
 gauge  fields  of  Yang-Mills or Maxwell type.  Characteristics of this
system of equations are studied for various gauge groups and the  liaison
between the new  and standard constructions of classical gauge fields
is discussed.
     Some ideas  proposed  in  [8, 9, 21] are used in the article.  Certain
elements of the construction proposed could be found in the  works  of
Hestenes [6], Casanova [30],
K\"ahler [19], Pestov [14], Pezzaglia and Differ [17]. A good
review of the techniques related to the Clifford algebra and a
comprehensive bibliography can be found in Keller's article [7, 23].
     Sign \fin\ will mark the end of the proof of a theorem  or  emphasize
the lack  of such proof which means that the proposition can be proved
by direct computation. In a number of cases the author used the
computer for analytic calculations. See also [31].

\subsection{Beyond the Standard Dirac Equation.}

Let us consider the standard Dirac equation and its generalizations.
 A vector\hfill\break
  $x=(x^0,x^1,x^2,x^3)\in \R^4$
defines a point in space-time, $x^0$ -- time coordinate,
$x^1,x^2,x^3$ -- space
coordinates and $\partial_\mu={\partial/{\partial x^\mu}},\ \mu=0,1,2,3$
are partial derivatives. A Klein-Gordon equation for
a function $\phi=\phi(x)$
$$
(\dal + m^2)\phi=0, \eqno(1.1)
$$
where $m$ -- nonnegative real number, describes particles with spin
$0$ and mass $m$. For a description of spin $1/2$ particles (electrons)
P.~A.~M.~Dirac suggested a system of equations of first order
that he obtained by formally factorizing the Klein-Gordon operator
$$
(i\gamma^\mu \partial_\mu + m)(i\gamma^\nu \partial_\nu - m)=-(\dal + m^2),
\eqno(1.2)
$$
where $\gamma^\mu,\ \mu=0,1,2,3$
are algebraic objects that satisfy the relations
$$
\gamma^\mu \gamma^\nu + \gamma^\nu \gamma^\mu = 2 g^{\mu\nu},
\quad \mu,\nu=0,1,2,3 \eqno(1.3)
$$
with Minkowski tensor $g=(g^{\mu\nu})={\rm diag}(1,-1,-1,-1)$.
$\gamma^\mu$ can be
faithfully represented by square matrices of order no
less than four. We shall use the following
``standard'' representation for the $\gamma^\mu$:
$$
\begin{array}{ccc}
&\gamma^0=\pmatrix{\sigma^0 & 0 \\ 0 & -\sigma^0}, \quad
\gamma^k=\pmatrix{0 & -\sigma^k \\ \sigma^k & 0}, \quad k=1,2,3 \\
&\sigma^0=\pmatrix{1&0\\ 0&1},\quad
\sigma^1=\pmatrix{0&1\\ 1&0},\quad
\sigma^2=\pmatrix{0&-i\\ i&0},\quad
\sigma^3=\pmatrix{1&0\\ 0&-1},
\end{array} \eqno(1.4)
$$
which is often called the Dirac representation ($\sigma^k$ --
Pauli matrices).
Let $\one$ denote the identity \4-matrix. The 16 matrices
$$
\begin{array}{ccc}
&\one; \qquad \gamma^\mu, \quad 0\leq\mu\leq3; \qquad
\gamma^{\mu\nu}\equiv\gamma^\mu \gamma^\nu, \quad 0\leq\mu<\nu\leq3; \\
&\gamma^{\mu\nu\lambda}\equiv\gamma^\mu \gamma^\nu \gamma^\lambda, \quad
0\leq\mu<\nu<\lambda\leq3; \qquad
\gamma^{0123}\equiv\gamma^0\gamma^1\gamma^2\gamma^3,
\end{array} \eqno(1.5)
$$
are linear independent and form a basis of ${\M}(4,{\C})$ -- algebra of four
dimensional complex matrices.

The Dirac equation has a form
$$
(i\gamma^\mu \partial_\mu-m \one)\psi=0, \eqno(1.6)
$$
and if we consider the aggregate in brackets as a \4-matrix, then
$\psi=\psi(x)$
 must be
a matrix with four lines and an arbitrary number of columns. The identity
(1.2) asserts that all the components of the matrix $\psi$ (if they are smooth
enough) satisfy the Klein-Gordon equation.

With the same result one can use the equation
$$
(i\gamma^\mu \partial_\mu+m \one)\psi^\prime=0. \eqno(1.7)
$$
If we multiply it from the left by $\gamma^{0123}$
 and define $\psi=\gamma^{0123}\psi^\prime$,  we obtain
(1.6) ($\gamma^{0123}$ anticommute with $\gamma^\mu$).

We shall call the equation (1.6) with one column matrix $\psi$
the standard Dirac equation. $\psi$ is called bispinor or Dirac spinor.
  One can also consider the
equations (1.6) with $l>1$ columns in matrix $\psi$. For  different
purposes physicists have used the Dirac equation (1.6) with
different numbers
of columns in $\psi$. The following list does not lay claim for
completeness
and indisputability:
\newpage

\begin{itemize}
\item[$l=1$:\ ] Spin $1/2$ particles in  quantum electrodynamics
(a gauge field theory with
$\U(1)$ symmetry group).
\item[$l=2$:\ ] Spin $1/2$ particles in  theory of electroweak
interactions ($\SU(2)$ gauge field theory).
\item[$l=3$:\ ] Spin $1/2$ particles in  theory of strong
interactions -- quantum chromodynamics
( $\SU(3)$ gauge field theory).
\item[$l=4$:\ ] The Dirac equation (1.6) with $l=4$ is
called a Rarita-Schwinger
equation. It is used for a description of spin $3/2$ particles.
\item[$l=5$:\ ] Georgi and Glashow [4] have suggested $\SU(5)$ gauge field
theory as Grand Unified Theory (GUT).
\item[$l\geq6$:\ ] $\SU(l)$
 gauge field theories are developed by physicists as
GUT.
\end{itemize}
\medskip

Let us note, that usually in theoretical physics people uses columns
(spinors) with $4l$ components,
but not $4\!\times\! l$-matrices, as wave functions
 (nevertheless, see Casanova [30]). One can easily
establish an equivalence between the Dirac equation for
$4l$-bispinor and the Dirac equation for $4\!\times\! l$-matrix.

There is an evident algebraic generalization of the identity (1.2)
$$
(i\gamma^\mu \partial_\mu +
m(z\one-y\gamma^{0123}))(i\gamma^\nu \partial_\nu -
m(z\one+y\gamma^{0123}))=-(\dal + m^2),
\eqno(1.8)
$$
where $z,y$ are complex constants and $z^2+y^2=1$. It leads to the equation
$$
(i\gamma^\mu \partial_\mu-m(z\one+y\gamma^{0123}))\psi=0, \eqno(1.9)
$$
We can consider the factorizations (1.2), (1.8) of
the Klein-Gordon operator as
one of many possible methods of a reduction of the Klein-Gordon
equation of second
order to a system of equations of first order (see Keller [7,
23]. There is another method
of such a reduction that leads to the following system of equations of
first order:
$$
i\gamma^\mu \partial_\mu \Psi-m( \Psi N + \gamma^{0123} \Psi K)=0, \eqno(1.10)
$$
\ti2 Theorem 1.
If the matrix $\Psi=\Psi(x)$ from ${\M}(4,{\C})$
 with twice continuously
differentiable
elements is a solution of (1.10), where the matrices $N,K\in{\M}(4,{\C})$
are not depend on $x$ and satisfy the relations
$$
[N,K]=NK-KN=0,\quad N^2+K^2=\one, \eqno(1.11)
$$
then the  matrix $\Psi$ is also a solution of the Klein-Gordon
equation.\par

\proof Let us consider an action of  the operator
$i\gamma^\mu \partial_\mu$ from left to the
equation (1.10)
$$
(i\gamma^\mu \partial_\mu)^2 \Psi - m (i\gamma^\mu \partial_\mu \Psi) N +
m\gamma^{0123}(i\gamma^\mu \partial_\mu \Psi) K=0
$$
and use the relations
$(i\gamma^\mu \partial_\mu)^2=-(\dal)$,
$i\gamma^\mu \partial_\mu \Psi=
m(\Psi N + \gamma^{0123}\Psi K)$ and (1.11).
As a result we get the Klein-Gordon equation $-(\dal+m^2)\Psi=0$\fin

The formula (1.10) gives us a set of equations that depend on two
matrices $N,K$ with the relations (1.11). How to describe all matrix
pairs $N,K$ that satisfy (1.11)? If we write the matrix $N$ using a Jordan
normal form, then we can find all the  corresponding matrices $K$ by doing
an elementary calculation [5]. Such a way leads to the 15 classes of pairs $N,K$
that depend on several parameters.
\medbreak
\xample 1. The matrices
$$
N=V\diag(z_1,z_2,z_3,z_4)V^{-1},\ \
K=V\diag(y_1,y_2,y_3,y_4)V^{-1},\eqno(1.12)
$$
where ${z_k}^2+{y_k}^2=1$, $V$-- nondegenerate matrix from $\M(4,\C)$,
 satisfy (1.11).
\medbreak
\xample 2. The matrices
$$
N=V\pmatrix{z & 1 & 0 & 0\cr
             0 & z & 1 & 0\cr
             0 & 0 & z & 1\cr
             0 & 0 & 0 & z} V^{-1},\quad
K=V\pmatrix{y & a & b & c\cr
             0 & y & a & b\cr
             0 & 0 & y & a\cr
             0 & 0 & 0 & y} V^{-1},
$$
where $z^2+y^2=1,\ y\neq0,\ a=-z/y,\ b=-1/(2y^3),\ c=-z/(2y^5)$
 and $V$-- nondegenerate matrix from $\M(4,\C)$, satisfy (1.11).
\medbreak
Let us consider a solution $\Psi$ of (1.10) with the matrices $N,K$
from the example 1 and connect it with the solutions of the standard Dirac
equation. The columns of the matrix $V$ denote by $v_k\ (k=1,2,3,4)$.
Then $v_k$ are the
eigen vectors of $N$ corresponding to the eigen values $z_k$.
Simultaneously, $v_k$ are
the eigen vectors of $K$ corresponding to the eigen values $t_k$. So, if we
multiply (1.10) from right by $v_k$ and denote $\psi_k=\Psi v_k$
then we come to four
equations
$$
i\gamma^\mu \partial_\mu\psi_k-m(z_k\one+y_k\gamma^{0123})\psi_k=0
$$
that have the form (1.9), or (1.6).

An equation (1.10), where $\Psi$ is a $4\times l$-matrix ($l=1,2,\ldots$)
and $N,K$ are $l\times l$-matrices, was studied in [21].
The Rarita-Schwinger equation can be considered as a special case
of  the equation (1.10) when $N=\one, K=0$.

Our aim in this paper
is to develop a gauge field theory for the equation (1.10).
In the next section we describe some mathematical tools
connected with the  Clifford algebra, which we shall use in our approach to
a gauge field theory.

\section{Clifford Algebra and Spinor Groups.}

In this section, for the sake of clarity and notation,
 we give necessary algebraic material which
otherwise is  known to
specialists (see, for example, [18, 20]). This makes the present
article almost self contained from the Clifford algebra point of
view.

\subsection{Clifford Algebra.}

Let ${\cal E}$ be an $n$-dimensional real
vector space with the basis vectors $e^1,\ldots,e^n$
 and $g=(g^{ij})_{i,j=1,\ldots,n}$ is a given real symmetric
nondegenerate matrix that we call metric (metric tensor) of the
vector space ${\cal E}$. Let us introduce a $2^n$-dimensional
real vector space $\V$, which contains ${\cal E}$, with the
basis consisting of $2^n$ vectors numbered by the ordered
multi-indices ($e$ is the scalar unit of the algebra)
$$
e,e^i,e^{i_1 i_2},\ldots,e^{1\ldots n},\quad\quad
1\leq i\leq n,\ 1\leq i_1< i_2\leq n,\ \ldots \eqno(2.1) $$
Let
us define a multiplication of vectors of $\V$ with the aid of
the following rules:
\medskip

\begin{itemize}
\item[1)]$ (\alpha U)V=U(\alpha V)=\alpha(U V)\quad \hbox{for}\quad
\forall U,V\in\V, \alpha\in\R $
\item[2)]$ (U+V)W=UW+VW,\quad  W(U+V)=WU+WV \quad  \hbox{for}\\
\forall U,V,W\in\V$
\item[3)]$ (UV)W=U(VW)\quad \hbox{for}\quad  \forall U,V,W\in\V$
\item[4)]$ eU=Ue=U\quad \hbox{for}\quad  \forall U\in\V$
\item[5)]$ e^i e^j+e^j e^i=2 g^{ij}e\quad \hbox{for}\quad
i,j=1,\ldots,n $
\item[6)]$ e^{i_1}\ldots e^{i_k}=e^{i_1\ldots i_k}\quad
\hbox{for}\quad 1\leq i_1 < \cdots < i_k\leq n.$
\end{itemize}
\medskip

Using these rules one can compute a result of multiplication of
arbitrary vectors of $\V$. The vector space $\V$ with the defined
multiplication of vectors is an associative algebra, that is called (a real)
Clifford algebra of the vector space ${\cal E}$ with the metric $g$ and
denoted by $\Gamma=\Gamma(g)$. If the matrix $g$ is diagonal with
$p$ pieces of $+1$ and $q$ pieces of $-1$ and $p+q=n$, then the
corresponding Clifford algebra is denoted by $\Gamma(p,q)$, and
by $\Gamma(n)$ if $q=0$. The basis vectors $e^{1}\ldots e^{n}$
are called the generators
of Clifford algebra. The elements of Clifford algebra $\Gamma$ are
called \gnumbers.
\medskip
\note  The Clifford algebra was introduced in the year 1878 by
the English
mathematician W. K. Clifford [15] in his analysis of the
Grassmann papers, making them accesible and more standard in
notation;  who called it {\sl geometrical
algebra}. We shall
use the term ``\g number" short of the term ``geometrical
number." The Dirac \g matrices (1.4) are, in particular, \g numbers.
\medskip

\g numbers $\sum_{i_1<\cdots<i_k} u_{i_1\ldots i_k}e^{i_1\ldots i_k}$
 are called  \g numbers of rank $k$; a \g number $u e$ of
rank 0 is
identified with the scalar $u$; a \g number $u_{1\ldots n}e^{1\ldots n}$
 of rank $n$ is
called a pseudoscalar. The sets of \g numbers of ranks $k=0,1,\ldots,n$
 are the
subspaces $\ss{k}\Gamma$ of $\Gamma$
$$
\Gamma=\ss{0}\Gamma\oplus \ss{1}\Gamma\oplus\cdots\oplus\ss{n}\Gamma=
\ddot{\Gamma}\oplus\dot\Gamma, \eqno(2.2)
$$
where $\ddot\Gamma=\ss{0}\Gamma\oplus\ss{2}\Gamma\oplus\ldots$,
$\dot\Gamma=\ss{1}\Gamma\oplus\ss{3}\Gamma\oplus\ldots$,
$\ss{1}\Gamma={\cal E}$. The dimension of the vector space $\ss{k}\Gamma$
equals the binomial
coefficient $C^k_n$ and  $\sum_{k=0}^n C_n^k=2^n$.
The dimensions of $\ddot\Gamma$ and $\dot\Gamma$ are equal to
$2^{n-1}$. Elements of $\ddot\Gamma$ are called even \g numbers and
elements of $\dot\Gamma$
 are called odd \g numbers. Sometimes it is convenient to
explicitly indicate a rank or an evenness of \g number in its notation --
$\ss{k}U\in\ss{k}\Gamma,\ \ddot U\in\ddot\Gamma,\ \dot U\in\dot\Gamma$.
If $U\in\Gamma$, then from (2.2) we get
$$
U=\sum_{k=0}^n \ss{k}U=\ddot U+\dot U.
$$
We shall use the projection operators $\ss{k}\pi:\Gamma\to\ss{k}\Gamma$
 such that $\ss{k}\pi(U)=\ss{k}U$.
A multiplication of two \g numbers of ranks $k$ and $s$ gives the sum of
\g numbers of ranks from $|k-s|$ to $p=\min(k+s,\ 2n-k-s)$ with increment 2
$$
\ss{k}U\ss{s}V=\ss{|k-s|}W+\ss{|k-s|+2}W+\cdots+\ss{p}W.
$$
Evidently
$\ddot U \ddot V, \dot U \dot V\in \ddot \Gamma;\quad
\ddot U\dot V, \dot U\ddot V\in\dot\Gamma$
and so, $\Gamma$ is $Z_2$-graded algebra. A set of even \g numbers
$\ddot\Gamma$ is a subalgebra of $\Gamma$ that is called the
even Clifford algebra.

An important property of the Clifford algebra $\Gamma$ is given by
the formula
$$
(u_i e^i)^2=(g^{ij}u_i u_j)e,
$$
that is the square of every vector from $\ss{1}\Gamma={\cal E}$
is a scalar. Usually the metric $g$ is diagonal
with $\pm1$ on the diagonal.

Let us define a conjugation of \g numbers using the following
relations: \hfill\break
$(e^{i_1}\ldots e^{i_k})^\star=e^{i_k}\ldots e^{i_1}$. Then
$$
(UV)^\star=V^\star U^\star,\quad U^{\star\star}=U\quad
\hbox{for}\quad\forall U,V\in\Gamma.
\eqno(2.3)
$$
then $(\ss{k}{U})^\star=(-1)^{k(k-1)\over2}\ss{k}{U}$.

Let ${\cal A}$ be a finite-dimensional algebra and $\Gamma_{\cal
A}$ be a tensor
product ${\cal A}\otimes\Gamma$. We can consider $\Gamma_{\cal
A}$ as a set of \g numbers of the following form
$$
u\otimes e+u_i\otimes e^i+\cdots+u_{1\ldots n}\otimes e^{1\ldots n}\equiv
u e_{{\cal A}}+u_i e_{{\cal A}}^i+\cdots+u_{1\ldots n}e_{{\cal A}}^{1\ldots n}
$$
with the coefficients $u,u_i,\ldots$
 from ${\cal A}$. In ${\cal A}\otimes\Gamma$ the basis
vectors of ${\cal A}$ commute with the basis vectors of
$\Gamma$, where we shall use
a convention that an index ${\cal A}$ in $u_{i_1\ldots
i_k}e_{{\cal A}}^{i_1\ldots i_k}$
 means that \g numbers belong to
$\Gamma_{\cal A}$
$$
u_{i_1\ldots i_k}\in{{\cal A}}\quad\hbox{and}\quad
u_{i_1\ldots i_k}e_{{\cal A}}^{i_1\ldots i_k}=
e_{{\cal A}}^{i_1\ldots i_k}u_{i_1\ldots i_k}=
u_{i_1\ldots i_k}\otimes e^{i_1\ldots i_k}.
$$
When ${\cal A}=\C$ -- the field of complex numbers, then an
index $\C$ in the basis
vectors notations may be omitted
$e^{i_1\ldots i_k}_\C=e^{i_1\ldots i_k}$
and the algebra $\Gamma_\C$ is called

a complex Clifford algebra.
A multiplication of the elements
of $\Gamma_{\cal A}$ is defined
$$
U V=(u_{i_1\ldots i_k}v_{j_1\ldots j_r})
\otimes(e^{i_1\ldots i_k}e^{j_1\ldots j_k}),\
U=u_{i_1\ldots i_k}e_{\cal A}^{i_1\ldots i_k},\
V=v_{j_1\ldots j_r}e_{\cal A}^{j_1\ldots j_r}.
$$
If ${\cal A}$ is an associative algebra with unit element, then
$\Gamma_{\cal A}$ is
also an associative algebra with unit element.

\note  In the case of a diagonal metric a complex Clifford algebra
$\gc(p,q)$ is trivially
isomorphic to $\gc(n)$. This is because the generators $e^k$
with the property $(e^k)^2=-e$ can be replaced by the generators
${\tilde e}^k=i e^k$.
\medskip

If $a\to \bar a$ is a conjugation operator in the algebra ${\cal
A}$ such that $\bar{ab} =\bar b \bar a,\ \bar{\bar a}=a$, then
we may introduce a  conjugation operator $U\to U^*$ in
$\Gamma_{\cal A}$ using the formula $U^*=\bar u_{i_1\ldots
i_k}e_{\cal A}^{i_k}\ldots e_{\cal A}^{i_1}$ for $U=u_{i_1\ldots
i_k}e_{\cal A}^{i_1\ldots i_k}$. Evidently $(UV)^*=V^* U^*$,\
$U^{**}=U$ for $U,V\in \Gamma_{\cal A}$ and $U^*=U^\star$ for
$U\in\Gamma$.

It should be pointed out that there is a possibility to define
the different conjugation operators $U\to T U^* T^{-1},\
U\in\Gamma_{\cal A}$ that depend on invertible \g number
$T\in\Gamma_{\cal A}$ with the property $T=T^*$ or $T=-T^*$.

Let us consider the properties of a transformation $V\to U V U^*$
that depend on some fixed $U\in\Gamma$.

\tit Theorem 1.

{\it \begin{itemize}
\item[1)]If $U\in\ss{1}\Gamma$, then $U\ss{1}\Gamma
U^*\subseteq\ss{1}\Gamma$.
\item[2)] If $U\in\ddot\Gamma$ and $n\leq 4$, then
$U\ss{1}\Gamma U^*\subseteq\ss{1}\Gamma$ and $UU^*=U^* U$.
\item[3)] If  $U\in\dot\Gamma$ and $n\leq 4$, then
$U\ss{1}\Gamma U^*\subseteq\ss{1}\Gamma$, but  \g numbers $U$
and $U^*$,
generally speaking, do not commute.
\item[4)] If $n\leq5,\ U\in\ddot\Gamma,\ U U^*=e $, then
$U\ss{1}{\Gamma}U^*\subseteq\ss{1}{\Gamma}$.
\item[5)] If $n\leq5,\ U\in\ddot\Gamma,\ U U^*=e $ and the metric
$g$ is diagonal, then
\end{itemize}}
$$
U\ss{k}{\Gamma}U^*\subseteq\ss{k}{\Gamma},\quad k=0,\ldots n\fin
$$

\note  There is a reason to suppose, that when $n=6$ and $g$ is a diagonal
metric, the conditions $U\in\ddot\Gamma(g),\  U U^*=e,\
U\ss{1}{\Gamma}U^*\subseteq\ss{1}{\Gamma}$ imply
$$
U\ss{k}{\Gamma}U^*\subseteq\ss{k}{\Gamma},\quad k=2,\ldots n.
$$
But we have not checked it.
\medbreak

\subsection{Spinor Groups.}

\df . The set of even \g numbers $F\in\ddot\Gamma$ such, that
$FF^*=e,\ F{\cal E} F^*\subseteq{\cal E}$ is a group (with respect to
multiplication)
that is called spinor group and denoted by $\Spin(g)$.

As a consequence of the proposition 4) of theorem 1 we get that for
$n\leq5$ group $\Spin(g)$ can be defined in a more simple way
$$
\Spin(g)=\{F\in\ddot\Gamma(g) : F F^*=e \}.
$$

For $F\in\Spin(g)$ the mapping ${\cal E}\to F{\cal E} F^*$
conserves the metric $g$
of the vector space ${\cal E}$. That means if $U=u_i e^i\in{\cal
E}$, then $V=FUF^*=v_i e^i\in{\cal E}$ and
$U^2=V^2=(g^{ij}u_i u_j)e=(g^{ij}v_i v_j)e$. This is a most important property
of spinor groups.

Let us consider in more details a case when the matrix $g$ is diagonal with
$p$ pieces of $+1$ and $q$ pieces of $-1$ on the diagonal and $p+q=n$.
A corresponding spinor group is denoted by $\Spin(p,q)$ and by $\Spin(n)$
 when $q=0$, As for $F\in\Spin(p,q)$ the mapping $\ss{1}U\to F\ss{1}U F^*$
of ${\cal E}$ into itself conserves the metric, then we get the homomorphism
$$
f: \Spin(p,q)\to O(p,q).
$$
The kernel of $f$ consists of $-1$ and $1$. The range of $f$ depends on the
numbers $p,q$: if $q=0$, then $O(p,q)=O(n)$ and $f(\Spin(n))=SO(n)$.
That means
$$
\Spin(n)/\{\pm1\}=SO(n).
$$
If $p>0, q>0$ then $f(\Spin(p,q))=SO^+(p,q)$, that means
$$
\Spin(p,q)/\{\pm1\}=SO^+(p,q),
$$
where $SO^+(p,q)$ is a connected component of unity element of the group
$SO(p,q)$.

We shall use the following theorem.

\ti2 Theorem 4 \hbox{[13]}.  The spinor groups
$$
\Spin(2),\ \Spin(3),\ \Spin(4),\ \Spin(5),\ \Spin(6)
$$
are isomorphic to the unitary groups
$$
\U(1),\ \Sp(1)\sim\SU(2),\ \SU(2)\times\SU(2),\ \Sp(2),\ \SU(4)
$$
respectively\fin\par

\subsection{Minkowski Metric.}

From this subsection on  we shall consider the case
$n=4$ and $g$-- Minkowski metric
$$
g=(g^{\mu\nu})_{\mu,\nu=0,1,2,3}={\rm diag}(1,-1,-1,-1).
$$
We shall use Greek indices $\mu,\nu,\ldots$ with values $0,1,2,3$.
${\cal E}$ is a four dimensional vector
space with the basis vectors $e^\mu$; $\Gamma=\Gamma(1,3)$ is a 16-dimensional
Clifford algebra; $\gc=\gc(1,3)$ is corresponding complex Clifford algebra.
An arbitrary \g number $U\in\Gamma,\ (U\in\gc)$ can be written as a linear
combination of the basis vectors of $\Gamma$ with the real (complex)
coefficients
$$
U=ue+u_\mu e^\mu+\sum_{\mu<\nu}u_{\mu\nu}e^{\mu\nu}+
\sum_{\mu<\nu<\lambda}u_{\mu\nu\lambda}e^{\mu\nu\lambda}+u_5 e^5,
\eqno(2.5)
$$
where $e^5=e^{0123},\ (e^5)^2=-e,\ u_5=u_{0123}$. The pseudoscalar $e^5$
commute with even \g numbers and anticommute with odd \g numbers.
Algebra $\Gamma$, as a vector space, can be decomposed into a direct sum of its
subspaces
$$
\Gamma=\ss{0}\Gamma\oplus
\ss{1}\Gamma\oplus\ss{2}\Gamma\oplus\ss{3}\Gamma\oplus\ss{4}\Gamma=
\ddot\Gamma\oplus\dot\Gamma, \quad\hbox{where}\quad
\ddot\Gamma=\ss{0}\Gamma\oplus
\ss{2}\Gamma\oplus\ss{4}\Gamma,\quad
\dot\Gamma=\ss{1}\Gamma\oplus\ss{3}\Gamma.
$$
The dimensions of the spaces
$\ss{0}\Gamma,\ss{1}\Gamma,\ss{2}\Gamma,\ss{3}\Gamma,\ss{4}\Gamma$
are equal to $1,4,6,4,1$ respectively, the dimensions of the spaces
$\ddot\Gamma$ and $\dot\Gamma$ are equal to $8$.

For $U=\sum_{k=0}^4 \ss{k}U \in\Gamma$ conjugated \g number has the form
$U^*=(\ss{0}U+\ss{1}U+\ss{4}U~)-(\ss{2}U+\ss{3}U~)$. Denoting
$\ss{014}\Gamma=\ss{0}\Gamma\oplus\ss{1}\Gamma\oplus\ss{4}\Gamma$,\
 $\ss{23}\Gamma=\ss{2}\Gamma\oplus\ss{3}\Gamma$, we get that $U=U^*$ for
 $U\in\ss{014}\Gamma$ and $U=-U^*$ for $U\in\ss{23}\Gamma$.

Let us consider the mapping of \g numbers $U\to F^* U F$ that depend on some
fixed \g number $F$.

\ti2 Theorem 2.  The following propositions are true for
$n=4$ and Minkowski metric:
\medskip
{\it
\begin{itemize}
\item[1)] If $F\in\Gamma$, then $F\ss{014}\Gamma
F^*\subseteq\ss{014}\Gamma,
\quad F\ss{23}\Gamma F^*\subseteq\ss{23}\Gamma$.
\item[2)] If $F\in\Gamma_\C$, then
$F\ss{014}\Gamma F^*\subseteq\ss{014}\Gamma\oplus i\ss{23}\Gamma,\
F\ss{23}\Gamma F^*\subseteq\ss{23}\Gamma\oplus i\ss{014}\Gamma$.
\item[3)] If $F\in\ddot\Gamma$, or $F\in\dot\Gamma$, then
$$
F(\ss{0}\Gamma\oplus\ss{4}\Gamma)F^*\subseteq\ss{0}\Gamma\oplus\ss{4}\Gamma,
\quad F\ss{k}\Gamma F^*\subseteq\ss{k}\Gamma, \quad k=1,2,3.
$$
\item[4)] If $F\in\ddot\Gamma_\C$, then
\end{itemize}
$$
\begin{array}{lll}
&F(\ss{0}\Gamma\oplus\ss{4}\Gamma)F^*\subseteq\ss{0}\Gamma\oplus\ss{4}
\Gamma\oplus i\ss{2}\Gamma, \quad
F\ss{1}\Gamma F^*\subseteq\ss{1}\Gamma\oplus i\ss{3}\Gamma,\cr
&F\ss{2}\Gamma F^*\subseteq\ss{2}\Gamma\oplus
i\ss{0}\Gamma\oplus i\ss{4}\Gamma,
\quad F\ss{3}\Gamma F^*\subseteq\ss{3}\Gamma\oplus i\ss{1}\Gamma\fin
\end{array}$$}
\par

\df. The mapping $U\to U^\dagger=e^0 U^* e^0$ is called a
hermitian conjugation of \g numbers.\par
\medbreak

This definition does not depend on matrix representations of $e^\mu$. But
nevertheless, the mapping $U\to U^\dagger$ in the matrix representation
(1.4) corresponds to the usual hermitian conjugation of matrices. This is
an advantage of the matrix representation (1.4). The matrix representations
$e^\mu\to{\tilde\gamma}^\mu=T^{-1}\gamma^\mu T$ taken from (1.4) with the
aid of unitary matrix $T$ have the same advantage.

\subsection{Trace, Determinant and Exponent of \g Numbers.}

From the matrix representation (1.4) we can immediately define
trace,
determinant and exponent of \g numbers. Important in Clifford
algebra is that  the same operators can
be defined independently of any matrix representations of the \g
numbers. This matrix independent definition of the
``determinant'' of \g numbers can be found, for example, in [29].

\df. The trace of a \g number $A\in\gc$ is a complex number
$\tr A=2\ss{0}{\pi}(A+A^\star)$.\par

The function $\tr:\Gamma_\C\to\C$ is linear
$\tr(\alpha A+\beta B)=\alpha\,\tr\,A+\beta\,\tr\,B,\ \alpha,\beta\in\C$ and
$\tr(e)=4$.

\df. The determinant of a \g number $A\in\gc$ is a complex number
$$
\det\,A=\det\,M(A), \eqno(2.6)
$$
where $M(A)$ is a matrix representation (1.4) of \g number $A$ and the
right hand part of (2.6) is the usual determinant of the matrix.\par

It is easily seen that  $\det(AB)=\det(A)\det(B),
\ A,B\in\Gamma_\C$ and $\det(\alpha A)=\alpha^4\det\,A,\ \alpha\in\C$,
$\det(e)=1$, and that with the aid of the determinant one can
introduce the notion of eigen-numbers
$\lambda_1,\lambda_2,\lambda_3,\lambda_4\in\C$ of a \g number
$A\in\Gamma_\C$
as the solutions of equation $\det(A-\lambda e)=0$, etc.

\df. The exponent of a \g number $A\in\gc$ is a \g number
$\exp\,A\in\Gamma_\C$ such that
$$
\exp\,A=e+\sum_{j=1}^\infty {1\over j!}A^j \eqno(2.7)
$$
\par
It is easy to check, that for an arbitrary \g number the series
(2.7) is convergent.

\ti2 Theorem 3.  Let $A\in\gc$. The function
$\exp:\Gamma_\C\to\Gamma_\C$ has the following properties:

\medskip
\begin{itemize}
\item[1)] $(\exp\,A)^*=\exp(A^*)$.
\item[2)] $(\exp\,A)^{\dagger} =\exp(A^\dagger)$.
\item[3)] If $V\in\Gamma_\C$ is invertible \g number, then
        $\exp(V^{-1}AV)=V^{-1}(\exp\,A)V$.
\item[4)] $\det(\exp\,A)=\exp(\tr\,A)$.
\item[5)] A \g number $\exp\,A$ is invertible and
$(\exp\,A)^{-1}=\exp(-A)$.
\item[6)] The mapping $\exp:\Gamma_\C\to\Gamma_\C$ is one to one
continuous
mapping of a small neighborhood of zero \g number on the small
neighborhood of identity \g number $e$.
\item[7)] If \g numbers $A,B$ commute, then
$(\exp\,A)(\exp\,B)=\exp(A+B)$.
\item[8)] If $A\in\Gamma_\C$ such that $A^2=-e$ and $\phi\in\R$,
then
$\exp(\phi A)=e\cos\phi+A\sin\phi$.
\end{itemize}
\medskip

\proof{\sl s} can be found in textbooks devoted to Lie groups
(see for example, Chevalley [12] or Cornwell [13])\fin

\subsection{Antihermitian Basis and Lie Algebras.}

For the complex Clifford algebra $\gc$ (see discussion in [28])
we shall use the basis
$$
ie^0,e^1,e^2,e^3,ie^{01},ie^{02},ie^{03},e^{12},e^{13},e^{23},e^{012},
e^{013},e^{023},ie^{123},e^5,ie \eqno(2.8)
$$
such that $t^\dagger=-t,\ t^2=-e and\ \tr(t^2)=-4$ and
where $t$ is an arbitrary basis vector. As $t^\dagger=-t$, the basis
(2.8) is called antihermitian basis of $\gc$ (the matrices (2.8)
in representation (1.4) are antihermitian).
We shall consider several Lie algebras and Lie groups that are
connected with the antihermitian basis.

Let $t_1,\ldots,t_n\ (n\leq16)$ be vectors from (2.8) which are
also generators of some real Lie algebra $\L$,
$$
[t_k,t_l]=c_{kl}^m t_m,\quad k,l=1,\ldots,n.
$$
where $c_{kl}^m$ are real valued structural constants of Lie algebra $\L$.
Sometimes we shall emphasize the dependence on the generators writing
$$
\L=\L(t_1,\ldots,t_n)=\{\tau_k t_k : \tau_k\in\R\},
$$

Let we consider real Lie algebras of dimensions 15,10,6,3 with generators
from (2.8). In particular
$$
\begin{array}{lll}
{\cal L}^{15}&={\cal L}^{15}(ie^0,e^1,e^2,e^3,ie^{01},ie^{02},ie^{03},
e^{12},e^{13},e^{23},e^{012},
e^{013},e^{023},ie^{123},e^5), \\
{\cal L}^{10}_1 &= {\cal L}^{10}_1(e^1,e^2,e^3,
e^{12},e^{13},e^{23},e^{012},
e^{013},e^{023},e^5), \\
{\cal L}^{10}_2 &= {\cal L}^{10}_2(ie^0,e^1,e^2,e^3,ie^{01},ie^{02},ie^{03},
e^{12},e^{13},e^{23}), \\
{\cal L}^{10}_3 &= {\cal L}^{10}_3(ie^{01},ie^{02},ie^{03},
e^{12},e^{13},e^{23},e^{012},
e^{013},e^{023},ie^{123}), \\
{\cal L}^{6}_1 &= {\cal L}^{6}_1(e^{12},e^{13},e^{23},e^{012},e^{013},e^{023}), \\
{\cal L}^{6}_2 &= {\cal L}^{6}_2(ie^{0},e^{1},e^{2},ie^{01},ie^{02},e^{12}), \\
{\cal L}^{6}_3 &= {\cal L}^{6}_3(ie^{01},ie^{02},ie^{03},e^{12},e^{13},e^{23}), \\
{\cal L}^{6^\prime}_1 &= {\cal L}^{6^\prime}_1(e^{12},e^{13},e^{23},ie^0,ie^{123},e^5), \\
{\cal L}^{3}_1 &= {\cal L}^{3}_1(e^{12},e^{13},e^{23}), \\
{\cal L}^{3}_2 &= {\cal L}^{3}_2(ie^{01},ie^{02},e^{12}).
\end{array}\eqno(2.9)
$$
The generators of $\L^{15}$ are all vectors of (2.8) with the exception
of $i e$. The generators of $\L^{10}_1$ are all real (without $i$) vectors
of (2.8). The generators of $\L^{10}_2,\L^{10}_3$  are the vectors from
(2.8) of ranks 1,2 and 2,3 respectively. The generators of $\L^6_1,
\L^6_2,\L^6_3$ are the vectors from (2.8) that commute with the \g numbers
$i e^0, e^{012}, e^5$ respectively. There are 15 Lie algebras  of the
dimension 6 with the generators from (2.8) that commute with one of the
basis vectors of (2.8) (with the exception of $i e$). Let us denote the
remaining Lie algebras by $\L^6_4,\ldots,\L^6_{15}$. The generators
of $\L^3_1,\L^3_2$ are even vectors that is also the generators of
$\L^6_1,\L^6_2$ respectively. To receive a Lie algebras of the dimension 3
one must take two arbitrary noncommuting vectors from (2.8) and
their product as a third generator. This method gives 20 Lie algebras
including $\L^3_1,\L^3_2$. Let us denote the remaining Lie algebras by
$\L^3_3,\ldots,\L^3_{20}$. The twenty Lie algebras
$\L^3_1,\ldots,\L^3_{20}$ can be divided into ten pairs
$\L^{6^\prime}_j=\L^3_{k_j} \oplus \L^3_{l_j}$, $j=1,\ldots,10$
such that if $u\in\L^3_{k_j}$, $v\in\L^3_{l_j}$, then $[u,v]=0$.
In particular
$\L^{6^\prime}_1=\L^3(e^{12},e^{13},e^{23}) \oplus \L^3(ie^0,ie^{123},e^5)$.
So we get 49 real semisimple Lie algebras
$$
\L^{15},\L^{10}_1,\L^{10}_2,\L^{10}_3,\L^6_1,\ldots,\L^6_{15},
\L^{6^\prime}_1,\ldots,\L^{6^\prime}_{10},
\L^3_1,\ldots,\L^3_{20} \eqno(2.10)
$$

\PROP. All Lie algebras from (2.10) of the same dimension (10, or 6,
or 3) are isomorphic.\par

\proof We have to consider generators and structure constants of
appropriate Lie algebras. For example, let us prove isomorphisms
$\L^3_1\sim\L^3_2$ and $\L^{6^\prime}_1\sim \L^6_3$.
Denoting $t_1=e^{12},\ t_2=e^{13},\ t_3=e^{23}$, we get relations
$$
[t_1,t_2]=2t_3,\quad [t_3,t_1]=2t_2,\quad [t_2,t_3]=2t_1\eqno(2.11)
$$
which can be written in a form
$[t_k,t_l]=2\epsilon_{klm} t_m$. Now, if one denote
$t_1^\prime=-ie^{01},\ t_2^\prime=-ie^{02},\ t_3^\prime=e^{12}$, then
\g numbers $t_1^\prime,t_2^\prime,t_3^\prime$ satisfy the same relations
(2.11) as $t_1,t_2,t_3$. This gives us isomorphism
$\L^3_1\sim\L^3_2$. By the way, it follows from (2.11), that these Lie
algebras are isomorphic to the Lie algebra
$\su(2)$ which consists of traceless antihermitian matrices of the second
order.

In order to prove isomorphism
$\L^{6^\prime}_1\sim \L^6_3$, we can denote
$$
\begin{array}{lll}
t_1&=(ie^{01}+e^{23})/2,\quad \hat t_1=(-ie^{01}+e^{23})/2,\cr
t_2&=(ie^{02}-e^{13})/2,\quad \hat t_2=(-ie^{02}-e^{13})/2,\cr
t_3&=(ie^{03}+e^{12})/2,\quad \hat t_3=(-ie^{03}+e^{12})/2.
\end{array}$$
and get
$$
[t_k,t_l]=2\epsilon_{klm}t_m,\quad
[\hat t_k,\hat t_l]=2\epsilon_{klm}\hat t_m,\quad
[t_k,\hat t_l]=0\quad k,l=1,2,3.
$$
That means, generators $t_1,t_2,t_3$ and $\hat t_1,\hat t_2,\hat t_3$
of the Lie algebra $\L^6_3$ play the same role as generators
$e^{12},e^{13},e^{23}$
and $ie^0,ie^{123},e^5$ of the Lie algebra $\L^{6^\prime}_1$, and hence,
$\L^6_3\sim\L^{6^\prime}_1$ (this also gives us isomorphism
$\Spin(4)\sim\SU(2)\times\SU(2)$)\fin
\medskip

Let $\L^{16}=\L^{16}(t_1,\ldots,t_{16})$ be a real Lie algebra with
16 generators from (2.8). Lie algebra
$\L^{16}=\L^1(i e)\oplus\L^{15}(i e^0,\ldots,e^5)$ is isomorphic to the
Lie algebra $u(4)=u(1)\oplus su(4)$ of all antihermitian matrices of the
dimension four. The following Lie algebras are the subalgebras of $u(4)$:
$$
\begin{array}{lll}
&\su(4),\quad\sp(2),\quad\su(3),\quad\su(2),\\
&\su(4)\oplus\u(1),
\quad\sp(2)\oplus\u(1),\quad\su(3)\oplus\u(1),\quad\su(2)\oplus\u(1),\\
&\su(2)\oplus\su(2),\quad\su(2)\oplus\su(2)\oplus\u(1),
\quad\su(2)\oplus\su(2)\oplus\u(1)\oplus\u(1),\\
&\su(2)\oplus\u(1)\oplus\u(1),\quad\su(2)\oplus\u(1)\oplus\u(1)\oplus\u(1),\\
&\u(1),\quad\u(1)\oplus\u(1),\quad\u(1)\oplus\u(1)\oplus\u(1),\quad
\u(1)\oplus\u(1)\oplus\u(1)\oplus\u(1).
\end{array}\eqno(2.12)
$$
The Lie algebras (2.12) are compact (their Killing forms are negatively
defined).

So, all of Lie algebras (2.10) are Lie subalgebras of
$\L^{16}\sim\u(4)\sim\u(1)\oplus\su(4)$
and there is a theorem.

\ti2 Theorem. The Lie algebras (2.10) are isomorphic to the
following classical matrix Lie algebras:
$$
\begin{array}{lll}
&\L^{15}\sim su(4),\cr
&\L^{10}_1,\L^{10}_2,\L^{10}_3\sim sp(2),\cr
&\L^6_1,\ldots,\L^6_{15},\L^{6^\prime}_1,\ldots,\L^{6^\prime}_{10}\sim
su(2)\oplus su(2),\cr
&\L^3_1,\ldots,\L^3_{20}\sim su(2)\fin
\end{array}$$
\par

\subsection{The Connection Between Complex Clifford Algebra and
Unitary Lie Groups.}

If ${\cal L}={\cal L}(t_1,\ldots,t_n),\ 1\leq n\leq16$ is a real
Lie algebra with
its generators from (2.8), then with the aid of exponential operator
we may introduce a Lie group
$$
{\cal G}={\cal G}(t_1,\ldots,t_n)=\{\exp(\tau_k t_k) : \tau_k\in\R\}.
$$
In this case Lie algebra ${\cal L}$ is a real Lie algebra of the linear Lie
group ${\cal G}$. Using the properties of the exponent:
$(\exp\,A)^\dagger=\exp(A^\dagger)$,
$(\exp\,A)^{-1}=\exp(-A)$ and the property
$t_k^\dagger=-t_k,\ k=1,\ldots,n$
we get $U^\dagger U=U U^\dagger=e$.
That means the Lie group ${\cal G}$ is unitary.

Let us write down some compact subgroups of the Lie group
$\U(4)$. In the first place, Lie groups
$\U(4),\U(3),\U(2),\U(1)$ are compact. Secondly, the following
Lie groups:
$$
\begin{array}{lll}
&\SU(4),\quad\Sp(2),\quad\SU(3),\quad\SU(2),\\
&\SU(4)\times\U(1),
\quad\Sp(2)\times\U(1),\quad\SU(3)\times\U(1),\quad\SU(2)\times\U(1),\\
&\SU(2)\times\SU(2),\quad\SU(2)\times\SU(2)\times\U(1),
\quad\SU(2)\times\SU(2)\times\U(1)\times\U(1),\\
&\SU(2)\times\U(1)\times\U(1),\quad\SU(2)\times\U(1)\times\U(1)\times\U(1),\\
&\U(1),\quad\U(1)\times\U(1),\quad\U(1)\times\U(1)\times\U(1),\quad
\U(1)\times\U(1)\times\U(1)\times\U(1).
\end{array}\eqno(2.12a)
$$
which can be derived from corresponding Lie algebras (2.12) with the aid of
exponential function, are compact. In a well known paper of Dynkin and
Oniscik [16] there is a method  of description of all compact subgroups of
the Lie group $\U(4)$.

By
$$
{\cal G}^{15},{\cal G}^{10}_1,{\cal G}^{10}_2,{\cal G}^{10}_3,{\cal G}^6_1,\ldots,{\cal G}^6_{15},
{\cal G}^{6^\prime}_1,\ldots,{\cal G}^{6^\prime}_{10},
{\cal G}^3_1,\ldots,{\cal G}^3_{20} \eqno(2.13)
$$
we denote linear semisimple Lie groups with the same generators as
corresponding real Lie algebras (2.10). In what follows we establish
isomorphisms between introduced Lie groups and classical matrix Lie groups.

First of all let us remind some known results on a Lie group theory. Here
$\Sp(n)$ is the unitary symplectic groups.

\ti2 Theorem 4. \hbox{[13]} The groups
$$
\Spin(2),\ \Spin(3),\ \Spin(4),\ \Spin(5),\ \Spin(6)
$$
are isomorphic to the groups
$$
\U(1),\ \Sp(1)\sim\SU(2),\ \SU(2)\times\SU(2),\ \Sp(2),\ \SU(4)
$$
and double cover groups
$$
\SO(2),\ \SO(3),\ \SO(4),\ \SO(5),\ \SO(6)
$$
respectively \fin\par

\ti2 Theorem 5. The Lie groups (2.13) are isomorphic to the following classical
Lie groups:
$$
\begin{array}{lll}
&{\cal G}^{15}\sim SU(4)\sim\Spin(6),\\
&{\cal G}^{10}_1,{\cal G}^{10}_2,{\cal G}^{10}_3\sim \Sp(2)\sim\Spin(5),\\
&{\cal G}^6_1,\ldots,{\cal G}^6_{15},{\cal G}^{6^\prime}_1,\ldots,{\cal G}^{6^\prime}_{10}\sim
SU(2)\times SU(2)\sim\Spin(4),\\
&{\cal G}^3_1,\ldots,{\cal G}^3_{20}\sim SU(2)\sim\Spin(3)\fin
\end{array}\eqno(2.14)
$$
\par

\proof It was proved in the proposition in the previous subsection, that
${\cal L}^3_1\sim \su(2)$, and hence
${\cal G}^3_1,\ldots,{\cal G}^3_{20}\sim\SU(2)\sim\Spin(3)$. So, if we prove
isomorphisms ${\cal G}^{15}\sim\SU(4),\ {\cal G}_1^{10}\sim\Spin(5),\ {\cal G}_3^6\sim\Spin(4)$,
then all the rest isomorphisms (2.14) will be follow from the theorem 4.

Let us begin with the isomorphism
${\cal G}^{15}\sim\SU(4)$. We can use Dirac's representation of
vectors of the antihermitian basis in the form of antihermitian
\4-matrices and get, that the real Lie algebra ${\cal L}^{15}$
is isomorphic to the Lie algebra $\su(4)$ of all traceless
antihermitian \4-matrices. This leads us to the conclusion, that
a linear Lie group ${\cal G}^{15}$ is isomorphic to the Lie
group $\SU(4)$ of special unitary matrices of the fourth order.
The fact that determinants of matrices is equal to $+1$ is a
consequence of the property $\det(\exp\,A)=\exp(\tr\,A)$.  So,
isomorphism ${\cal G}^{15}\sim\SU(4)$ is proved.

Let us prove the isomorphism ${\cal G}^{10}_1\sim\Spin(5)$. We may denote
generators of the group ${\cal G}^{10}_1$ as
$$
\{t_1,\ldots,t_{10}\}=\{e^1,e^2,e^3,
e^{12},e^{13},e^{23},e^{012},e^{013},e^{023},e^{0123}\}\in\Gamma(1,3).
$$
Let us consider a Clifford algebra
$\Gamma(5)$ with generators $f^1,\ldots,f^5$ which satisfy the relations
$f^k f^l+f^l f^k=2 \delta^{kl} f,\ k,l=1,\ldots,5$, where $f$ is an identity
vector. We may define \g numbers
$\hat t_1,\ldots,\hat t_{10}$ by the formula
$$\begin{array}{lll}
\{\hat t_1,\ldots,\hat t_{10}\}=\cr
\{-f^{14},-f^{24},-f^{34},
-f^{12},-f^{13},-f^{23},f^{35},-f^{25},f^{15},-f^{45}\}
\in\ss{2}{\Gamma}(5)
\end{array}
$$
where $f^{kl}=f^k f^l$ for $k<l$. It is easy to check, that (in all
formulas the summation convention is used)
$$
[\hat t_k,\hat t_l]=
c^m_{kl}\hat t_m,\quad k,l=1,\ldots,10,\quad
c^m_{kl}\in\R \eqno(2.15)
$$
and hence, $\hat t_k$ are generators of a real Lie algebra
$$
\hat{\cal L}^{10}=\hat{\cal L}^{10}(\hat t_1,\ldots,\hat t_{10})=
\{\tau_k\hat t_k : \tau_k\in\R\}
$$
and of a linear Lie group
$$
\hat{\cal G}^{10}=\hat{\cal G}^{10}(\hat t_1,\ldots,\hat t_{10})=
\{\exp(\tau_k\hat t_k) : \tau_k\in\R\}.
$$
Since $\hat t_k\in\ss{2}{\Gamma}(5)$, then $\hat t_k^*=-\hat t_k$.
If $\hat U=\exp(\tau_k\hat t_k)\in\hat{\cal G}^{10}$, then, using the properties
of exponent
$(\exp\,A)^*=\exp(A^*),\ (\exp\,A)^{-1}=\exp(-A)$, we get that
$\hat U \hat U^*=e$. By the theorem 1 of section 2 we have
$\hat U\ss{1}{\Gamma}(5)\hat U^*\subseteq\ss{1}{\Gamma}(5)$.
Therefore, the linear Lie group
$\hat{\cal G}^{10}$ is isomorphic to the group $\Spin(5)$.

At last we may check (it can be done by a direct calculation), that the
generators $t_1,\ldots,t_{10}$ of the group ${\cal G}^{10}_1$ have the same
structure constants
$[t_k,t_l]=c^m_{kl} t_m,\ k,l=1,\ldots,10$ as the generators
$\hat t_k$ in (2.15). This leads us to isomorphism
${\cal G}^{10}_1\sim\hat{\cal G}^{10}\sim\Spin(5)$.

Finally, let us prove the isomorphism ${\cal G}^6_3\sim\Spin(4)$. Generators of
the Lie group ${\cal G}^6_3$ are \g numbers
$$
\{t_1,\ldots,t_6\}=\{ie^{01},ie^{02},ie^{03},e^{12},e^{13},e^{23}\}.
$$
Let us denote $e^4=ie^0$. In that case
$e^k e^l+e^l e^k=2\delta^{kl}e,\ k,l=1,2,3,4$ and
$$
\{t_1,\ldots,t_6\}=\{e^{kl}\}_{k<l\leq4} \eqno(2.16)
$$
Hence, a Lie algebra ${\cal L}^6_3$ is isomorphic to $\ss{2}{\Gamma}(4)$
-- a set of \g numbers of the second rank of the real Clifford algebra
$\Gamma(4)$. Let us show, that a Lie group
${\cal G}^6_3$ is isomorphic to the Lie group $\Spin(4)$. From the relation
(2.16) we have $t_k^*=-t_k$ and so, for all $U\in{\cal G}^6_3$ there is an
equality $U U^*=e$. By the theorem 1 of section 2 we get
$U\ss{1}{\Gamma}(4) U^*\subseteq\ss{1}{\Gamma}(4)$. Therefore, by the
definition of spinor groups, our Lie group
${\cal G}^6_3$ is isomorphic to the group $\Spin(4)$.
The theorem is proved\fin\medskip

\consequence 1. The real Lie algebras of the Lie groups $\Spin(n),\ n=2,3,4,5$
are isomorphic to the Lie algebras $\ss{2}{\Gamma}(n)$ with respect
to commutator $[A,B]=A B - B A$\fin\par

I don't know whether this proposition is true for Lie groups
$\Spin(n),\ n>5$.

\consequence 2. Among  49 Lie groups (2.13) the Lie groups of
the dimensions 15, 10, 3 are simple and the Lie groups of the
dimension 6 are semisimple.\par

Let us formulate several additional theorems, which will be
useful later on.

\ti2 Theorem \hbox{[13]}. If ${\cal G}$ is a linear Lie group and ${\cal
L}$ is its real Lie algebra, then for arbitrary $U\in{\cal G},\
v\in{\cal L}$ ($U,v$ may smoothly depend on  $x\in\R^4$), the
aggregates $U^{-1}vU,\ U^{-1}\partial_\mu U$ belong to ${\cal
L}$\fin\par

Let ${\cal G}$ be a linear Lie group and ${\cal L}={\cal
L}(t_1,\ldots,t_n)$ its real Lie algebra. For every $U\in{\cal
G}$ let $\Ad(U)$ is a real $n\!\times\!n$-matrix that is defined
by the relations
$$
U t_k U^{-1}=\{\Ad(U)\}_{jk} t_j,\quad k=1,\ldots,n.\eqno(2.17)
$$

\ti2 Theorem 7 \hbox{[13]}. The set of real matrices $\Ad(U)$ is
$n$-dimensional representation of the linear Lie group ${\cal
G}$ that is called adjoint representation\fin\par

From (2.17) it is evident that the matrices of adjoint
representation are depend on the generators $t_1,\ldots,t_n$.
For the unitary Lie group ${\cal G}$ every matrix $\Ad(U)$ is
similar to the orthogonal matrix $\Ad(U)=T^{-1}OT$, $
O^T=O^{-1}$, $T\in\M(n,\R)$.

\df. The generators $t_1,\ldots,t_n$ of a Lie algebra ${\cal L}$
of the unitary Lie group ${\cal G}$ are said to satisfy ADRIO
condition (ADjoint Representation Is Orthogonal), if for every
$U\in{\cal G}$ the matrix $\Ad(U)$ is orthogonal.\par

\ti2 Theorem 8. If the generators  $t_1,\ldots,t_n$ of a real Lie
algebra ${\cal L}$ of the unitary Lie group ${\cal G}$ satisfy
ADRIO condition, then for every $U\in{\cal G}$

$$
\begin{array}{lll}
U t_k U^\dagger&=\{\Ad(U)\}_{lk} t_l,\quad k=1,\ldots,n\\
U^\dagger t_l U&=\{\Ad(U)\}_{lk} t_k,\quad l=1,\ldots,n.
\end{array}\eqno(2.18)
$$
\par
\proof The first equality in (2.18) is a definition of an adjoint
representation. From it, replacing
$U\leftrightarrow U^\dagger$,
$k\leftrightarrow l$, we get
$$
U^\dagger t_l U=\{\Ad(U^\dagger)\}_{kl} t_k,\quad l=1,\ldots,n.\eqno(2.19)
$$
For the adjoint representation we have
$\Ad(U^\dagger)=\Ad(U^{-1})=(\Ad U)^{-1}$. ADRIO condition gives
$(\Ad U)^{-1}=(\Ad U)^T$ and hence\hfill\break
$\{\Ad(U^\dagger)\}_{kl}=\{\Ad U\}_{lk}$. By substituting this to (2.19)
we get the second equality in (2.18)\fin

Let ${\cal G}^{15}$ be a linear Lie group from (2.13) with the generators from
(2.8).

\ti2 Theorem 9. The generators (2.8) of the Lie group ${\cal G}^{15}\sim\SU(4)$
satisfy ADRIO condition\fin\par

As a consequence of this theorem we get that the generators of 49 Lie
groups (2.13) satisfy ADRIO condition.

\ti2 Theorem. Let ${\cal G}$ be a linear Lie group, ${\cal L}$
-- its real Lie algebra
of the dimension $n$, $t_1,\ldots,t_n$ and
$\tau_1,\ldots,\tau_n$  --- two
sets of generators of the Lie algebra
${\cal L}$ which connected by relations
$t_k=h_{lk}\tau_l$, where $h_{lk}$ -- elements of an orthogonal
$n\!\times\!n$-matrix $H$. For every $U\in{\cal G}$ let
$\Ad_t(U)=(a_{lk})$ and $\Ad_\tau(U)=(\alpha_{lk})$ be
$n\!\times\!n$-matrices of an adjoint representation of the group
${\cal G}$ corresponding to generators
$t_k$ and $\tau_k$
$$
U t_k U^{-1} = a_{jk}t_j,\quad U\tau_k U^{-1}=\alpha_{jk}\tau_j,\quad
k=1,\ldots,n. \eqno(2.19a)
$$
Then, matrices $\Ad_t(U)$ and $\Ad_\tau(U)$ are connected by the relation
$$
\Ad_\tau(U)=H(\Ad_t(U)) H^T. \eqno(2.19b)
$$
\par

\proof Substituting
$t_k=h_{lk}\tau_l$, $t_j=h_{mj}\tau_m$ (summation convention is used) into
the first equality in (2.19a), we get
$$
U h_{lk}\tau_l U^{-1}=a_{jk} h_{mj}\tau_m.
$$
Let us multiply on $h_{ik}$ both parts of this equality and sum with
respect to index $k$
$$
U h_{ik}h_{lk}\tau_l U^{-1}=a_{jk} h_{mj}h_{ik}\tau_m.
$$
Using the orthogonality property of the matrix $H$ ---
$h_{ik}h_{lk}=\delta_{il}$, we come to the relation
$U\tau_i U^{-1}=a_{jk}h_{mj}h_{ik}\tau_m$.
Comparing it with the second equality in
 (2.19a), we get
$\alpha_{mi}=a_{jk}h_{mj}h_{ik}$ which equivalent to
(2.19b)\fin\par
\medskip
A following theorem is a consequence of the proved theorem.

\ti2 Theorem 10. Let ${\cal G}$ be a unitary Lie group and ${\cal L}$ its real Lie algebra
of the dimension $n$ with generators
$t_1,\ldots,t_n$ which satisfy ADRIO condition. And let
$\tau_1,\ldots,\tau_n$ be another set of generators of
${\cal L}$ such, that
$t_k=h_{lk}\tau_l$ where $h_{lk}$ -- elements of an orthogonal matrix $H$.
Then generators
$\tau_1,\ldots,\tau_n$ also satisfy ADRIO condition\fin\par

\subsection{Gell-Mann's Generators of the Lie Group $\SU(3)$.}

Let ${\cal G}$  be a Lie group
$\U(4)$ or $\U(1)\times\SU(4)$, and ${\cal L}=u(1)\oplus su(4)$ be a real Lie
algebra of ${\cal G}$. Vectors of antihermitian basis
(2.8) are generators of the Lie algebra ${\cal L}$ that satisfy ADRIO condition.
Let us introduce a new generators
$$
\begin{array}{lll}
t_{1}&=( e^{023} +  e^{23})/ \sqrt{2}\\
t_{2}&=(-{{ e^{013}} + { e^{13}}})/ {{\sqrt{2}}}\\
t_{3}&=({{ e^{012}} + { e^{12}}})/ {{\sqrt{2}}}\\
t_{4}&=({{ e^{0123}} - { e^{03}} i})/ {{\sqrt{2}}}\\
t_{5}&=(-{{ e^{3}} + { e^{123}} i})/ {{\sqrt{2}}}\\
t_{6}&=(-{{ e^{2}} + { e^{01}} i})/ {{\sqrt{2}}}\\
t_{7}&=(-{{ e^{1}} - { e^{02}} i})/ {{\sqrt{2}}}\\
t_{8}&=({ e^{012} -  e^{12} - 2  e^{0} i})/ {{\sqrt{6}}}\\
t_{9}&=({{ e^{2}} - { e^{01}} i})/ {{\sqrt{2}}}\\
t_{10}&=(-{{ e^{1}} + { e^{02}} i})/ {{\sqrt{2}}}\\
t_{11}&=({{ e^{0123}} + { e^{03}} i})/ {{\sqrt{2}}}\\
t_{12}&=({{ e^{3}} - { e^{123}} i})/ {{\sqrt{2}}}\\
t_{13}&=(- e^{023} +  e^{23})/ {{\sqrt{2}}}\\
t_{14}&=({{ e^{013}} - { e^{13}}})/ {{\sqrt{2}}}\\
t_{15}&=(- e^{012} +  e^{12} -  e^{0} i)/ {{\sqrt{3}}}\\
t_{16}&=ie
\end{array}\eqno(2.20)
$$
which are expressed from generators (2.8) with the aid of orthogonal
$16\!\times\!16$-matrix and $\tr({t_k}^2)=-4$ as for generators
(2.8). Vectors (2.20) can be considered as a new basis of complex
Clifford's algebra $\gc$. By the theorem 10, generators (2.20)
satisfy ADRIO condition. In representation (1.4) matrices
$t_1,\ldots,t_8$ from (2.20) have a following structure:
$3\!\times\!3$-matrices in the left upper corner of
$t_1,\ldots,t_8$ are equal to Gell-Mann's matrices
$i\sqrt2\lambda_1,\ldots,i\sqrt2\lambda_8$ which are conventional
generators of
the Lie group
$\SU(3)$. The fourth columns and the fourth lines in the matrices
$t_1,\ldots,t_8$ are composed from zeros. Hence, the generators
$t_1,\ldots,t_8$ from (2.20) of a Lie group ${\cal G}^8(t_1,\ldots,t_8)\sim\SU(3)$
satisfy  ADRIO condition.

\subsection{Spinorial Generators of the Lie Group
$\U(1)\times\U(1)\times\U(1)\times\U(1)$.}

Let us consider an Abelian Lie
group
${\cal G}=\U(1)\times\U(1)\times\U(1)\times\U(1)$ isomorphic to the group of
diagonal matrices from
$\M(4,\C)$ with absolute values of all diagonal elements equal to one.
The Lie group ${\cal G}$ is maximal Abelian subgroup of a Lie group
$\U(4)$. A real Lie algebra
${\cal L}=\u(1)\oplus\u(1)\oplus\u(1)\oplus\u(1)$
of the Lie group $G$ is isomorphic to an algebra of diagonal matrices from
$\M(4,\C)$ with pure imaginary elements on the diagonal. As the generator
of ${\cal L}$ we can take vectors $ie,ie^0,e^{12},e^{012}$ of the
antihermitian
basis (2.8), that are diagonal matrices in representation
 (1.4). Let us introduce new generators of the Lie algebra ${\cal L}$:
$$\begin{array}{lll}
t_1&=(ie+ie^0-e^{12}-e^{012})/2,\\
t_2&=(ie+ie^0+e^{12}+e^{012})/2,\\
t_3&=(ie-ie^0-e^{12}+e^{012})/2,\\
t_4&=(ie-ie^0+e^{12}-e^{012})/2
\end{array}\eqno(2.21)
$$
which are expressed from $ie,ie^0,e^{12},e^{012}$ with the aid of
orthogonal \4-matrix, and
$\tr({t_k}^2)=-4$ as for (2.8) and
(2.20). In the representation (1.4) generators $t_k$ are diagonal
\4-matrices with only nonzero element
$(t_k)_{kk}=2i$. Generators
(2.21) are called spinorial generators of the Lie algebra
$\u(1)\oplus\u(1)\oplus\u(1)\oplus\u(1)\subset\u(4)$
(of the Lie group
$\U(1)\times\U(1)\times\U(1)\times\U(1)\subset\U(4)$).  By the
theorem 10, spinorial generators (2.21) satisfy ADRIO condition.
This concludes our presentation of the algebraic background.

\section{The Dirac \g Equation.}

Let $g$ be Minkowski metric, $\gc=\gc(1,3)$, the Clifford
complex algebra
$x=(x^0,x^1,x^2,x^3)\in\R^4$, $x^\mu$ -- coordinates of a point
in space-time, ${\cal D}^k(\R^4,\Gamma_\C)$  be a space of
$k$-times continuously differentiable functions that map $\R^4$
into $\gc$ (all coefficients of the \g number $U\in{\cal
D}^k(\R^4,\Gamma_\C)$ are $k$ times continuously differentiable
functions of $x\in\R^4$). We shall use the operator
$\dsl=e^\mu\partial_\mu$ such that $\dsl^2=e(\dal)$. If
$U\in{\cal D}^1(\R^4,\Gamma_\C)$, then the partial derivatives
$\partial_\mu U$ are \g numbers with the coefficients that are
partial derivatives of the corresponding coefficients of $U$.

The main equation, that is called a Dirac \g equation has the form
$$
i\dsl\Psi-m(\Psi N+e^5 \Psi K)=0, \eqno(3.1)
$$
where $\Psi\in{\cal D}^1(\R^4,\Gamma_\C)$, $m$-- real number,
the commuting
$N,K\in\gc$ do not depend on $x$ and satisfy the decomposition
condition  (1.11).  If
$\Psi\in{\cal D}^2(\R^4,\Gamma_\C)$ is a solution of (3.1),
then, as it was proved in theorem 1 of sec. 1, \g number $\Psi$
is also a solution of the Klein-Gordon equation (1.1).

In this paper we do not study the generalized solutions of (3.1) (that
belong to one or another space of generalized functions). In what follows
a record like  $\Psi=\Psi(x)\in\gc$ means that \g number $\Psi$ has
coefficients that are smooth functions of $x\in\R^4$.

If we multiply the equation (3.1) on $-i$ and denote
$\hat N=i N,\ \hat K= i K$, then we come to an equation
$$
\dsl\Psi+m(\Psi \hat N+e^5 \Psi \hat K)=0, \eqno(3.1a)
$$
where
$$
{\hat N}^2+{\hat K}^2=-e,\quad [\hat N, \hat K]=0. \eqno(3.1b)
$$
Such \g-numbers $\hat N, \hat K$ can be from $\Gamma_\C$, or
from $\Gamma$. The equation (3.1a) in which
$\hat N, \hat K\in\Gamma$ satisfy (3.1b) and
$\Psi\in{\cal D}^k(\X,\Gamma),\ k\geq1$ is called a real Dirac
\g-equation.

\subsection{Decomposition of the Dirac \g Equation Into Even and
Odd Parts.}

\ti2 Theorem 2. If $\Psi\equiv\ddot\Psi+\dot\Psi$ is a solution of
(3.1) in which $N,K\in\dot\Gamma_\C$,
then $\ddot\Psi,\dot\Psi$ are the solutions of the equations
$$
i\dsl\ddot\Psi-m\ddot\Psi N_0=0, \eqno(3.2)
$$
$$
i\dsl\dot\Psi-m\dot\Psi N_1=0. \eqno(3.3)
$$
where $N_0=N+e^5 K,\ N_1=N-e^5 K$ and ${N_0}^2={N_1}^2=e$.
So, (3.1) is decomposed into two independent equations (3.2) and (3.3).
\par
{\bf Proof} is evident\fin

The real Dirac \g-equation (3.1a) where
$\hat N, \hat K\in\dot\Gamma$ satisfy (3.1b) and
$\Psi\in{\cal D}^k(\X,\Gamma),\ k\geq1$ is decomposed into two equaions
$$
\dsl\ddot\Psi+m\ddot\Psi \hat N_0=0, \quad
\dsl\dot\Psi+m\dot\Psi \hat N_1=0. \eqno(3.4)
$$
where $\hat N_0=\hat N+e^5 \hat K,\ \hat N_1=\hat N-e^5 \hat K$ and
${\hat N_0}^2={\hat N_1}^2=-e$.
It is easy to calculate that a general form of \g-number $\hat N_0$
is the following:
$$
\hat N_0 = q_\mu e^\mu (\alpha e+\beta e^5),
$$
where real numbers $\alpha,\beta,q_0,\ldots,q_3$ satisfy the relation
$$
(\alpha^2+\beta^2)({q_0}^2-{q_1}^2-{q_2}^2-{q_3}^2)=-1.
$$
In particular, if we take
$\alpha=q_0=q_1=q_2=0,\ \beta=q_3=1$,
then we come to an equation
$$
\dsl\ddot\Psi+m\ddot\Psi e^{012}=0,
$$
which is called Hestenes equation [6, 22]. This equation was investigated
in many papers see references in [23]).

\subsection{Plane Wave Solutions.}

In quantum mechanics the plane wave solutions of the Dirac
equation are used for the construction of wave packets that
describe the fermion dynamics. The equation (3.1) has a more
rich set of plane wave solutions (3.6) in comparison with the
standard Dirac equation.

Let $\com(N,K)$ be an algebra of \g numbers from $\gc$ that
commute with $N,K$ and
$$
\S=\{S\in\com(N,K) : S^2=-e\}.
$$

\ti2 Theorem 1. The equation (3.1) with the conditions (1.11) has solutions
of the form
$$
\Psi=(\ss{1}P+im(N-e^5 K)S)\exp(p\!\cdot\! x S)Y,\eqno(3.5)
$$
where $S\in\S$, $Y\in\com(N,K)$ are \g numbers independent of $x$,
$\ss{1}P=p_\mu e^\mu,\ p\!\cdot\! x = p_\mu x^\mu$ and the real numbers
$p_\mu$ satisfy the relations $p_\mu p^\mu=m^2$.
\par

\proof In the case $Y=e$ a direct calculation gives
$$
e^\mu \partial_\mu\Psi=\ss{1}P\Psi S, \quad
im(\Psi N+e^5\Psi K)= -\ss{1}P\Psi S\fin
$$
\par
Hence, the plane wave solutions (3.5) depend on some real numbers $p_\mu$
that satisfy the relation $p_\mu p^\mu=m^2$ and they also depend on
\g numbers $S,Y$. The dependence of $\Psi$ on $Y$ , as will be shown, is
a consequence of a (global) gauge invariance of the equation (3.1).
Let us consider in more detail a dependence of (3.5) on $S\in\S$.
Let $S_k\in\S,\ k=1,\ldots,n$ be such \g numbers, that the sets
$$
\S_k=\{U^{-1} S_k U : U\in\com(N,K),\ \det U\neq0\}
$$
cover the set $\S$, so $\S=\S_1\cup\ldots\cup\S_n$.
Then (3.5) gives the following set of solutions of (3.1):
$$
\Psi_k(p_\mu,U,Y)=
(\ss{1}P+im(N-e^5 K)U^{-1} S_k U)\exp(p\!\cdot\! x U^{-1}S_k U) Y,\eqno(3.6)
$$
where $U,Y\in\com(N,K),\ \det U\neq0$.
If $S_k$ belong to the center of the algebra $\com(N,K)$, then
the dependence on $U$ disappears from the formula (3.6).
In particular, the \g numbers
$e,N,K,N^a K^b$ ($a,b$ -- integer positive numbers)  and their linear
combinations belong to the center of $\com(N,K)$.

As examples, let us consider plane wave solutions of three equations.
We begin with the equation
i$\dsl\Psi-m\Psi=0$, where $N=e,K=0$. Evidently, $\com(N,K)=\gc$,
$\S=\{S\in\gc : S^2=-e\}$. The set $\S$ can be considered as a set of matrices
$\S=\{S\in\M(4,\C) : S^2=-{\bf1}\}$.
Each matrix $S\in\M(4,\C)$ with the aid of the similarity transformation
$S\to S^\prime=U^{-1}SU$, where $U\in\M(4,\C),\ \det U\neq0$ can be
transformed to a normal Jordan form  $S^\prime$. The condition
$S^2=-{\bf1}$ gives $(S^\prime)^2=-{\bf1}$. So, there are no Jordan cells
in $S^\prime$ and
$$
S^\prime=i\diag(\pm1,\pm1,\pm1,\pm1).\eqno(3.6a)
$$
Evidently, there are five matrices of the form (3.6a) that are not similar to
each other
$$
S_{1,2}=\pm i{\bf1},\ \ S_3=i\diag(1,1,-1,-1),\ \ S_{4,5}=\pm i\diag(1,-1,-1,-1).
$$
Hence, (3.6) gives five sets of plane waves, two of which correspond to
$S_1,S_2$ and do not depend on  $U$, because $S_1,S_2$ belong to the center
of $\gc$. We can write  $S_1,\ldots,S_5$ as a linear combinations of basis
vectors (2.8)
$$
S_{1,2}=\pm ie,\ \ S_3=ie^0,\ \ S_{4,5}=\pm i(-e+e^0+ie^{12}+ie^{012})/2.
$$

Now, let us consider plane wave solutions of the equation
$$
\dsl\Psi+m\Psi e^0 i=0,
$$
in which $\Psi\in{\cal D}^k(\X,\ddot\Gamma_\C)$. Substituting
$S=-i,\ K=0,\ N=e^0,\ Y=a e^0$ into  (3.5), we get
$$
\Psi=a(\ss{1}{P}+m e^0)e^0\exp(-p\cdot x i). \eqno(3.6b)
$$
If we take $a=1/\sqrt{2m(E+m)}$, where
$E=p^0=\sqrt{{p_1}^2+{p_2}^2+{p_3}^2+m^2}$, then the solution (3.6b)
satisfy the relation $\Psi \Psi^*=e$. In matrix representation (1.4)
$\Psi$ has a form:
$$
\Psi=\sqrt{{E+m}\over{2m}}\pmatrix{
1&0&{-p_3}\over{E+m}&{-p_1+i p_2}\over{E+m}\cr
0&1&{-p_1-i p_2}\over{E+m}&{p_3}\over{E+m}\cr
{-p_3}\over{E+m}&{-p_1+i p_2}\over{E+m}&1&0\cr
{-p_1-i p_2}\over{E+m}&{p_3}\over{E+m}&0&1} \exp(-p\cdot x i).
$$
Compare this formula with the formula (3.7) from a textbook [24].

Finally, let us consider the plane wave solutions of the equation
$$
\dsl\Psi+m\Psi e^0 I=0, \eqno(3.6c)
$$
where $\Psi\in{\cal D}^k(\X,\ddot\Gamma)$, and \g-number
$I=(q_1e^1+q_2e^2+q_3e^3)e^5$
depends on three real numbers $q_1,q_2,q_3$, which satisfy the relation
${q_1}^2+{q_2}^2+{q_3}^2=1$. Let us note, that $I^2=-e,\ [I,e^0]=0$.
The equation (3.6c) has a solution
$$
\Psi={1\over\sqrt{2m(E+m)}}(\ss{1}{P}+m e^0)e^0\exp(-p\cdot x I),
$$
which can be get from (3.6b) replacing $i$ by $I$. If $I=e^{12}$, the
last formula gives us a solution of Hestenes equation.

\subsection{Gauge Groups of the Dirac \g Equation.}

The Dirac \g equation (3.1) is invariant with respect to the (global)
transformation
$\Psi\to\Psi^\prime=\Psi U$, where $U\in\com(N,K)$
and $U$ is independent on $x$. It follows from the equality
$$
i\dsl\Psi^\prime-m\Psi^\prime N-me^5\Psi^\prime K=
(i\dsl\Psi-m\Psi N-me^5\Psi K)U,\eqno(3.6a)
$$
which is violated when $U=U(x)$, because in the expression
$\partial_\mu(\Psi U)=(\partial_\mu\Psi)U+\Psi\partial_\mu U$
there is a new term $\Psi\partial_\mu U$.
If we want to get a local invariance of the equation (3.1) under the
transformation  $\Psi\to\Psi U$ with $U=U(x)$, then the term
$\Psi\partial_\mu U$ must be compensated by something. The problem can be
solved by gauging of the equation (3.1). It consists of replacing of all
partial derivatives $\partial_\mu\Psi$ by so called covariant derivatives
$D_\mu\Psi=\partial_\mu\Psi-\Psi a_\mu$, where $a_\mu=a_\mu(x)\in\Gamma_\C$
are values with the following transformation rule
$a_\mu\to U^{-1}a_\mu U+U^{-1}\partial_\mu U$.

Let us introduce a unitary group
$$
{\cal G}_{max}(N,K)=\{U\in\Gamma_\C : [U,N]=[U,K]=0,\ U U^\dagger=e\}=
\com(N,K)\cap\U(4),
\eqno(3.7)
$$
which is called a maximal gauge group of the equation (3.1). The group
${\cal G}_{max}(N,K)$ is a compact Lie group.

\df. If a Lie group ${\cal G}$ is a Lie subgroup of
${\cal G}_{max}(N,K)$, then the group ${\cal G}$ is called a
gauge group of the Dirac \g equation (3.1). A real Lie algebra
${\cal L}$ of the group ${\cal G}$ is called a gauge Lie algebra
of the equation (3.1).\par

\df. If ${\cal G}$ is a gauge group of (3.1) and ${\cal L}$ its
real Lie algebra, then an equation $$ i
e^\mu(\partial_\mu\Psi-\Psi a_\mu)-m(\Psi N + e^5\Psi K)=0
\eqno(3.8) $$ where $a_\mu=a_\mu(x)\in{\cal L}$, is called a
Dirac \g equation with a gauge field $a_\mu$. The transformation
$$
\begin{array}{lll}\Psi&\to\Psi^\prime=\Psi U,\\
a_\mu&\to a_\mu^\prime=U^{-1}a_\mu U+U^{-1}\partial_\mu U
\end{array} \eqno(3.9)
$$
is called a gauge transformation of the equation  (3.8). \par

It should be noted, that the value
$ a_\mu^\prime=U^{-1}a_\mu U+U^{-1}\partial_\mu U$ belongs to the gauge Lie
algebra of (3.8) (theorem 6, sec.2).

\ti2 Theorem 3. The Dirac \g equation with a gauge field
$a_\mu\in{\cal L}$ is invariant under the gauge transformation (3.9) (in other
words, the equation (3.8) is invariant under the gauge group ${\cal G}$).

\proof It is a consequence of the equality
$$
ie^\mu(\partial_\mu\Psi^\prime-\Psi^\prime a_\mu^\prime)-m(\Psi^\prime N+
e^5\Psi^\prime K)=
(ie^\mu(\partial_\mu\Psi-\Psi a_\mu)-m(\Psi N+e^5\Psi K))U\fin\eqno(3.10)
$$

Let us remind, that to each Lie group ${\cal G}$ corresponds a unique real Lie
algebra ${\cal L}$. But for fixed real Lie algebra ${\cal L}$ there exists, generally
speaking, a class of Lie groups $\tilde{\cal G}$, such that ${\cal L}$ is a real Lie
algebra for every Lie group from this class. Lie groups $\tilde{\cal G}$
are locally isomorphic (homeomophic), but not isomorphic. For example,
the Lie algebra $\u(1)\oplus\su(4)$ is a real Lie algebra of the Lie group
$U(4)$ and also of the Lie group
$\U(1)\times\SU(4)$, This two Lie groups are  not isomorphic.

In many cases for us it will be suitable to deal with the subclass of
gauge groups, that uniquely defined by their real Lie algebras.

\df. If a real Lie algebra
 ${\cal L}={\cal L}(t_1,\ldots,t_n)=
\{\tau_k t_k : \tau_k\in\R\}$ with the antihermitian generators
$t_k^\dagger=-t_k,\ t_k\in\gc$ is a gauge Lie algebra of the equation  (3.1),
then the unitary Lie group
${\cal G}={\cal G}(t_1,\ldots,t_k)=\{\exp(\tau_k t_k) : \tau_k\in\R\}$ is called an
exponential gauge group of the equation (3.1).\par

All the exponential groups of (3.1) are compact Lie groups. For the real
Lie algebras (2.12) the correspondent exponential Lie groups were listed
in (2.12a).

\subsection{Conservation Laws.}

\df. Let
$j^\mu=j^\mu(\Psi),\ \mu=0,1,2,3$ be real functions that depend on a
solution $\Psi$ of the equation (3.1). If there is an equality
$\partial_\mu j^\mu=
\partial_0 j_0-\partial_1 j_1-\partial_2 j_2-\partial_3 j_3=0$,
then it is  called a conservation law of the equation (3.1) (written in
differential form).
In that case the vector $j=(j^0,j^1,j^2,j^3)$ is called a current of the
equation (3.1).
\par
Let us denote $\bar\Psi=\Psi^\dagger e^0=e^0 \Psi^*$.

\ti2 Theorem 4. Let $\Psi=\Psi(x)\in\Gamma_\C$ be a solution of the equation
(3.1) in which \g numbers $N,K$ are such, that
$$
\begin{array}{lll}
N&=\alpha_1 e + \beta_1 P_1,\quad P_1 P_1^\dagger=e,\quad P_1^\dagger=P_1,
\quad \alpha_1,\beta_1\in{\cal R}\\
K&=\alpha_2 e + \beta_2 P_2,\quad P_2 P_2^\dagger=e,\quad P_2^\dagger=P_2,
\quad \alpha_2,\beta_2\in{\cal R}
\end{array}\eqno(3.11)
$$
and let ${\cal L}={\cal L}(t_1,\ldots,t_n)\subseteq\Gamma_\C$ be a gauge Lie algebra of
the equation (3.1) with generators
$t_1,\ldots,t_n$ from the antihermitian basis (2.8),
or from Gell-Mann's basis (2.20), or from spinorial basis (2.21).
Let us define values
$j^\mu_k=j^\mu_k(\Psi)\in\R,\ \mu=0,1,2,3$; $k=1,\ldots,n$ by the formula
$$
\ss{1}\pi(\Psi t_k \bar\Psi i)=-g_{\mu\nu}j^\mu_k e^\nu.
\eqno(3.12)
$$
In that case there are equalities $\partial_\mu j^\mu_k=0,\ k=1,\ldots,n$,
which mean that vectors
$j_k=(j^0_k,j^1_k,j^2_k,j^3_k)$ are currents of the Dirac \g equation
 (3.1).\par

To prove this theorem we have to use three lemmas.

\LEM 1. If $\Psi\in{\cal D}^1(\X,\Gamma_\C)$ is a solution of the equation (3.1),
then it satisfies the relations
$$
i\partial_\mu(\bar\Psi e^\mu \Psi)-
m(\bar\Psi\Psi N-N^\dagger \bar\Psi\Psi+\bar\Psi e^5\Psi K-
K^\dagger \bar\Psi e^5\Psi)=0,\eqno(3.13)
$$
$$
i\partial_\mu(\bar\Psi e^5e^\mu \Psi)-m(\bar\Psi e^5\Psi N+
N^\dagger \bar\Psi e^5\Psi-
\bar\Psi\Psi K-K^\dagger \bar\Psi\Psi)=0.\eqno(3.14)
$$
\par
\proof In order to prove the relation  (3.13) we may multiply by
\g number $\bar\Psi$ an equation (3.1) from left and subtract a hermitian
conjugated equation
$-i\partial_\mu\Psi^\dagger  (e^\mu)^\dagger - m(N^\dagger \Psi^\dagger -
K^\dagger \Psi^\dagger e^5)=0$, that was multiplied from right on
$e^0\Psi$. The result can be written in the form (3.13).
In the same manner we can prove the relation (3.14). \fin

\LEM 2. Let ${\cal L}$ be a real Lie algebra of the gauge Lie group
${\cal G}$ of the Dirac \g equation (3.1), and \g numbers $N,K$ in the
equation (3.1) satisfy conditions (3.11). Then
$$
\ss{{\cal L}}\pi(\bar\Psi\Psi N-N^\dagger \bar\Psi\Psi+\bar\Psi \gamma^5\Psi K-
K^\dagger \bar\Psi \gamma^5\Psi)=0,\eqno(3.15)
$$
where $\ss{{\cal L}}\pi:\Gamma_\C\to{\cal L}$ is a projector operator to
the algebra
 ${\cal L}$ being  considered as a subspace of the linear space
 $\Gamma_\C$.

\proof
Let us denote
$$
B_1=\bar\Psi\Psi N-N^\dagger \bar\Psi\Psi,\quad
B_2=\bar\Psi e^5\Psi K-K^\dagger \bar\Psi e^5\Psi.
$$
These \g numbers $B_1,B_2$ anticommute with the \g numbers $P_1,P_2$
respectively:
$$
B_1 P_1+P_1 B_1=0,\quad B_2 P_2+P_2 B_2=0,
$$
that may be checked by the direct calculation.
The last fact can be written in the form:
$$
B_1\in\Gamma_\C \setminus  \com(P_1),\quad
B_2\in\Gamma_\C \setminus  \com(P_2),\eqno(3.16)
$$
where $\com(P)\subseteq\Gamma_\C$ is a subspace of \g numbers that commute
with the \g number $P\in\Gamma_\C$.
It follows from (3.16), that
$$
B_1+B_2\in\gc\setminus (\com(P_1)\cap\com(P_2))\eqno(3.17)
$$
Using the relations $\com(P_1)=\com(N)\ $, $\com(P_2)=\com(K)$ and by
definition of a gauge group
 ${\cal G}$, we get from (3.17), that
$\ss{{\cal L}}\pi(B_1+B_2)=0$\fin

\LEM 3. Let $\Psi=\Psi(x)\in\gc$ and
${\cal L}={\cal L}(t_1,\ldots,t_n)\subseteq\gc$ be an arbitrary real Lie algebra with
generators $t_1,\ldots,t_n$  from antihermitian basis or from Gell-Mann's
basis. If we define values
$j^\mu_k=j^\mu_k(\Psi),\ $, $\mu=0,1,2,3\ $;
$k=1,\ldots,n$ by the formula (3.12), then
$$
\ss{{\cal L}}\pi(\bar\Psi i e^\mu\Psi)=j^\mu_k t_k.\eqno(3.18)
$$
\par
\proof The proof of this lemma can be done by calculation for the Lie
algebra
 ${\cal L}=\gc\sim\su(4)\oplus\u(1)$ with generators from the antihermitian basis
 and from Gell-Mann's basis. This give us also a proof of lemma 3 for all
 subalgebras of ${\cal L}$
with generators from the antihermitian basis and Gell-Mann's basis.
Also, we may prove lemma 3 by calculation for the Lie algebra
${\cal L}=\u(1)\oplus\u(1)\oplus\u(1)\oplus\u(1)$ with spinorial generators
\fin

\medskip
{\bf Proof} of the theorem. The \g number $\Psi=\Psi(x)\in\gc$
satisfy the Dirac \g equation and so, by lemma 1,
$\Psi$ satisfy also a relation (3.13), and by lemma2, we get
$$
\ss{{\cal L}}\pi(\partial_\mu(\bar\Psi e^\mu\Psi))=0. \eqno(3.19)
$$
Let us define values $j^\mu_k=j^\mu_k(\Psi)\in\R$ by the formula (3.12).
By lemma 3, these values satisfy relations (3.18) and hence
$$
\partial_\mu(\ss{{\cal L}}\pi(\bar\Psi i e^\mu\Psi))=\partial_\mu j^\mu_k t_k,
$$
which together with (3.19) give
$$
\partial_\mu j^\mu_k t_k=0\fin
$$

\subsection{Canonical Forms of the Dirac \g Equation.}

In what follows, we consider equation (3.1) where \g numbers $N,K$ satisfy
conditions (1.11) and (3.11) simultaneously.

\ti2 Theorem 5. \g numbers $N,K$ satisfy conditions (1.11) and (3.11) if and
only if in representation (1.4) matrices $N,K$ have a form
$$
N=\cos\xi\,\one, \quad K=\sin\xi\,\one, \eqno(3.20)
$$
or
$$
\begin{array}{lll}
N&=U^\dagger\diag(\cos\xi,\cos\xi,\cos\xi,-\cos\xi)U,\\
K&=U^\dagger\diag(\sin\xi,\sin\xi,-\sin\xi,-\sin\xi)U,
\end{array}\eqno(3.20a)
$$
or
$$
\begin{array}{lll}
N&=U^\dagger\diag(\cos\xi,\cos\xi,\cos\xi,\cos\eta)U,\\
K&=U^\dagger\diag(\sin\xi,\sin\xi,\sin\xi,\sin\eta)U,
\end{array}\eqno(3.21)
$$
or
$$
\begin{array}{lll}
N&=U^\dagger\diag(\cos\xi,\cos\xi,\cos\eta,\cos\eta)U,\\
K&=U^\dagger\diag(\sin\xi,\sin\xi,\sin\eta,\sin\eta)U,
\end{array}\eqno(3.22)
$$
where $U$ -- unitary matrix and $0\leq \xi,\eta<2\pi$.
\par
\proof From the condition (3.11)
$N=(\alpha_1 \one + \beta_1 P_1)$, $ \alpha_1,\beta_1\in\R$,
$P_1=P_1^\dagger$, ${P_1}^2=\one$, that means $P_1$ is a hermitian and
simultaneously unitary matrix. Hence, one can reduce it to the diagonal
form with the aid of unitary matrix
 $U$:
$$
P_1=U^\dagger\diag(\pm1,\pm1,\pm1,\pm1)U,
$$
and
$$
N=U^\dagger\diag(\lambda_1,\lambda_2,\lambda_3,\lambda_4)U,
$$
where $\lambda_k=\alpha_1 \pm \beta_1$.
The same is true for the matrix $K$:
$$
K=V^\dagger\diag(\epsilon_1,\epsilon_2,\epsilon_3,\epsilon_4)V,
$$
where $V^\dagger V=\one$, $\epsilon_k=\alpha_2 \pm \beta_2$.
Matrices $N,K$ are commute by the conditions (1.11) and therefore they can
be reduced to the diagonal form by the same similarity transformation [5].
Hence, we can take $U=V$. A condition $N^2+K^2=\one$ gives
${\lambda_k}^2+{\epsilon_k}^2=1$. Therefore $\lambda_k=\pm\cos\xi$,
$\epsilon_k=\pm\sin\xi$, or
 $\lambda_k=\cos\phi_k$, $\epsilon_k=\sin\phi_k$
and $\phi_k$ may have no more than two values \fin
\bigbreak

If \g numbers $N,K$ in representation (1.4) have a form (3.20), or
(3.20a), or (3.21), or(3.22) with $U=\one$, then we say that the Dirac
\g equation (3.1) is written in a canonical form. So, there are four
canonical forms of the Dirac \g equation. Let us write correspondent \g
numbers $N,K$ with the aid of the basis elements of Clifford algebra
$$
N=\cos\xi\,e,\quad K=\sin\xi\, e, \eqno(3.23)
$$
or
$$
N=\cos\xi(e+e^0+i e^{12}-i e^{012})/2,\quad K=\sin\xi e^0, \eqno(3.23a)
$$
or
$$
\begin{array}{lll}
N&={\cos\xi\over4} (3e+e^0+ie^{12}-ie^{012})+
  {\cos\eta\over4} (e-e^0-ie^{12}+ie^{012}),\\
K&={\sin\xi\over4} (3e+e^0+ie^{12}-ie^{012})+
  {\sin\eta\over4} (e-e^0-ie^{12}+ie^{012}),
\end{array}\eqno(3.24)
$$
or
$$
\begin{array}{lll}
N&={\cos\xi\over2} (e+e^0)+{\cos\eta\over2} (e-e^0),\\
K&={\sin\xi\over2} (e+e^0)+{\sin\eta\over2} (e-e^0).
\end{array}\eqno(3.25)
$$
Every canonical form of the Dirac \g equation has its own gauge group
(we shall write only exponential gauge groups).
Four canonical forms of the Dirac \g equation
(3.23), (3.23a), (3.24), (3.25) correspond to the following gauge groups:
${\cal G}=\U(1)\times\SU(4)$,
${\cal G}=\U(1)\times\U(1)\times\SU(2)$,
${\cal G}=\U(1)\times\SU(3)$,
${\cal G}=\U(1)\times\U(1)\times\SU(2)\times\SU(2)$ respectively.
As it was shown in the theorem 2, section 3, if
 $N,K\in\dot\gc$, then the Dirac \g equation decompose into two independent
 equations for
$\ddot\Psi$ and $\dot\Psi$. Considering formulas
(3.23),(3.23a),
(3.24),(3.25) one can easily conclude that \g numbers $N,K$ can be odd
only for the fourth canonical form (3.25), when
$\eta=\xi\pm\pi$, that means $N=\cos\xi\,e^0$, $K=\sin\xi\,e^0$.
In that case we get the following variant of the Dirac \g equation:
$$
ie^\mu\partial_\mu\Psi-m\,\exp(\xi e^5)\Psi e^0=0,\ \xi\in\R,\ \Psi\in\ddot\gc,
\ \hbox{or}\ \Psi\in\dot\gc,
\eqno(3.26)
$$
with the gauge group ${\cal G}=\U(1)\times\SU(2)$. This equation, probably, can
be used in the theory of electroweak interactions. The first or the third
canonical form, probably, can be used in the theory of strong interactions
(quantum chromodynamics). In Grand Unified Theory, probably,
the first canonical form of the Dirac \g equation  can be used
(see the remark at the end of paper).

It is suitable to have ready examples of the simplest equations to
which the present theory can be applied. Such examples can be
derived from
(3.23), (3.24), (3.25), (3.26) when
$\xi=0,\ \eta=\pi$.
Let us write them down together with their gauge groups:
$$
\begin{array}{lll}
&i\dsl\Psi- m\Psi=0,\ \ \Psi\in\gc,\ \ {\cal G}=\U(1)\times\SU(4),\cr
&i\dsl\Psi-m(\Psi N + e^5 \Psi K)=0,\ \ \Psi\in\gc,\
\ {\cal G}=\U(1)\times\SU(3),\cr
&i\dsl\Psi- m\Psi e^0=0,\ \ \Psi\in\gc,\ \ {\cal G}=\U(1)\times\U(1)\times\SU(2)\times\SU(2),\cr
&i\dsl\Psi- m\Psi e^0=0,\ \ \Psi\in\ddot{\Gamma}_\C,\ \
{\cal G}=\U(1)\times\SU(2)
\end{array}$$
where in the second equation \g numbers $N,K$ satisfy relations (3.24).

\section{The Yang-Mills and Maxwell \g Equations.}
Let ${\cal G}$ be a linear Lie group and
${\cal L}={\cal L}(t_1,\ldots,t_n)$ its real Lie algebra. Let us consider Lie algebra
valued functions  $a_\mu=a_\mu(x)=a_\mu^k(x) t_k\in{\cal L}$ which depend on
$x=(x^0,x^1,x^2,x^3)\in\R^4$. We may also consider a rank 1 \g number
$A=a_\mu e_{\cal L}^\mu\in{\cal L}\otimes\ss{1}{\Gamma}$ and rank 2 \g number
$F=\sum_{\mu<\nu} f_{\mu\nu} e^{\mu\nu}_{{\cal L}}\in{\cal L}\otimes\ss{2}{\Gamma}$
with coefficients from the Lie algebra  ${\cal L}$
($f_{\mu\nu}=f_{\mu\nu}^k(x) t_k$).

\df. The system of equation
$$
\begin{array}{lll}
&\ss{2}\pi(e^\mu_{\cal L}(\partial_\mu A-[A,a_\mu])-A^2)-F=0,\\
&e^\mu_{\cal L}(\partial_\mu F-[F,a_\mu])=0,
\end{array}\eqno(4.1)
$$
is called the Yang-Mills \g equations (special form of the Yang-Mills
equations) with a gauge group ${\cal G}$.\par

A left hand part of the first equation is a rank 2 \g number with
coefficients from  ${\cal L}$. A left hand part of the second equation is a sum
of rank 1 \g number and rank 3 \g number. That means the second equation
from (4.1) can be written in the form of two identities
$$
\begin{array}{lll}
&\ss{1}{\pi}(e^\mu_{\cal L}(\partial_\mu F-[F,a_\mu]))=0,\\
&\ss{3}{\pi}(e^\mu_{\cal L}(\partial_\mu F-[F,a_\mu]))=0.
\end{array}\eqno(4.2)
$$
It is easy to check that the second identity in (4.2) is fulfilled for
arbitrary
$F\in\ss{2}{\Gamma}_{\!{\cal L}},\ a_\mu\in{\cal L}$ which depend smoothly on $x$.
This is a so-called Bianci identity. By substituting
$A=a_\mu e_{\cal L}^\mu,\ F=\sum_{\mu<\nu} f_{\mu\nu} e^{\mu\nu}_{\cal =
L}$ into the
first equation (4.1) and (4.2) and by equating to zero appropriate
coefficients of \g numbers, we get a standard Yang-Mills equations
$$
\begin{array}{lll}
&\partial_\mu a_\nu-\partial_\nu a_\mu-[a_\nu,a_\mu]-f_{\mu\nu}=0,\\
&\partial^\mu f_{\mu\nu}-[f_{\mu\nu}, a^\mu]=0,
\end{array}\eqno(4.3)
$$
where $\partial^\mu=g^{\mu\nu}\partial_\nu,\ a^\mu=g^{\mu\nu} a_\nu$ and
$f_{\mu\nu}=-f_{\nu\mu}$ for $\mu\geq \nu$.

If a Lie algebra ${\cal L}$ is of dimension 1 Abelian with the generator $t_1=i$,
then the commutators in (4.3) are equal to zero and we get the standard
Maxwell equations
$$
\begin{array}{lll}
&\partial_\mu a_\nu-\partial_\nu a_\mu-f_{\mu\nu}=0,\\
&\partial^\mu f_{\mu\nu}=0,
\end{array}\eqno(4.4)
$$
where $a_\mu=a_\mu^1 i,\ f_{\mu\nu}=f_{\mu\nu}^1 i$.
Now, if we define $A=a_\mu e^\mu,\ F=\sum_{\mu<\nu}f_{\mu\nu} e^{\mu\nu}$,
then Maxwell's equations (4.4) can be written in the form
$$
\begin{array}{lll}
&\ss{2}\pi(e^\mu\partial_\mu A)-F=0,\\
&e^\mu\partial_\mu F=0,
\end{array}\eqno(4.5)
$$
\df. A system of equations (4.5) is called the Maxwell \g equations (a
special form of Maxwell's equations).\par

\note  If we take an  equation
$e^\mu\partial_\mu A-F=0$ instead of the first equation in (4.5), then it
decomposes into two identities -- one of which is a first equation from
(4.5), and the second can be written in a form
$\partial^\mu a_\mu=0$ and called a Lorentz condition.

\subsection{A Gauge Invariance of the Yang-Mills and\hfill\break
 Maxwell \g Equations.}
\df. A local transformation (it depends on
$x\in\R^4$)
$$
\begin{array}{lll}
& a_\mu\to a_\mu^\prime=U^{-1}a_\mu U+U^{-1}\partial_\mu U, \\
& f_{\mu\nu}\to f_{\mu\nu}^\prime=U^{-1} f_{\mu\nu}U,
\end{array}\eqno(4.6)
$$
where $U=U(x)\in{\cal G},\quad\mu,\nu=0,1,2,3$ is called a gauge transformation
of the system of equations (4.1). Here a Lie group ${\cal G}$ and its real Lie
algebra ${\cal L}$ is called a gauge Lie group and a gauge Lie algebra of the
system (4.1) respectively.\par

Note, that by the theorem 6, section 2 we get
 $a_\mu^\prime,\ f_{\mu\nu}^\prime \in{\cal L}$.

\ti2 Theorem 1. A system of equations (4.1) is invariant under the gauge
transformation (4.6).\par

\proof An invariance of the system of equations (4.1) under a gauge
transformation means, that if
$a_\mu,\ f_{\mu\nu}$
$(A=a_\mu e^\mu_{\cal L},\ F=\sum_{\mu<\nu} f_{\mu\nu} e^{\mu\nu})$ satisfy
equations (4.1), then the functions
$a_\mu^\prime,\ f_{\mu\nu}^\prime,\ (A^\prime,\ F^\prime)$ from
 (4.6)  must also satisfy equations (4.1).
Hence, to prove a gauge invariance it is sufficient to substitute
$a_\mu^\prime,\ f_{\mu\nu}^\prime$ to the left hand parts of the equations
(4.1) and check the validity of the equations
$$
\begin{array}{lll}
\ss{2}\pi(e^\mu_{\cal L}(\partial_\mu
A^\prime-[A^\prime,a_\mu^\prime])-(A^\prime)^2)-F^\prime=\\
&\hbox{\hskip-3cm} =U^{-1}\{\ss{2}\pi(e^\mu_{\cal L}(\partial_\mu
A-[A,a_\mu])-A^2)-F\}U,\\
e^\mu_{\cal L}(\partial_\mu F^\prime-[F^\prime,a_\mu^\prime])&
\hbox{\hskip-3cm} =
U^{-1}\{e^\mu_{\cal L}(\partial_\mu F-[F,a_\mu])\}U\fin
\end{array}\eqno(4.7)
$$

In the case of Abelian Lie group
$$
{\cal G}=\U(1)=\{\exp\, i\lambda: \lambda\in\R\}
$$
Maxwell's \g  equations (4.5) are invariant under the gauge transformation
$$
\begin{array}{lll}
& a_\mu\to a_\mu^\prime=a_\mu + i \partial_\mu \lambda, \\
& f_{\mu\nu}\to f_{\mu\nu}^\prime= f_{\mu\nu},
\end{array}\eqno(4.8)
$$
which  can be got from (4.6) when $U=\exp\, i\lambda(x)$.

\section{A Construction of the Gauge Fields Theory.}

In this section we introduce a main system of equations, named a
Dirac-Yang-Mills system of \g equations, which describes Dirac's field
interacting with the Yang-Mills gauge field.

\subsection{A Union of the Dirac \g Equation and the\hfill\break
 Yang-Mills \g Equations.}

Let us consider an equation (3.1) and associate with it a commutator
algebra $\com(N,K)$ of \g numbers from $\gc$ which commute with $N$ and $K$
simultaneously. A set of invertible \g numbers from $\com(N,K)$ is a linear
Lie group ${\cal G}_0={\cal G}_0(N,K)$. Let a Lie group ${\cal G}$ be an arbitrary Lie
subgroup of
${\cal G}_0$ and ${\cal L}$ be a real Lie algebra of ${\cal G}$ with generators
$t_1,\ldots,t_n\ (n\leq16)$. And let  $J$ be an operator
$J:\gc\to{\ss{1}{\Gamma}}_{\cal L}$ with the following properties:
\medskip
1) $J(\Psi U)=U^{-1}J(\Psi)U$ for all $U\in{\cal G}$.
\medskip
2) $J(\Psi)=j_\mu^k(\Psi) t_k\otimes e^\mu$, where functionals
$j^k_\mu:\gc\to\R$ are such that if $\Psi$ satisfy an equation (3.1), then
$\partial^\mu j^k_\mu(\Psi)=0,\ k=1,\ldots,n$, that means,
$j_k=(j_k^0,\ldots,j_k^3)$ are the currents of the equation (3.1).
\medskip

In that case we can consider a system of equations for
 $\Psi\in\gc$,
$A=a_\mu e^\mu_{\cal L}\in{\ss{1}{\Gamma}}_{\cal L}$, $F\in{\ss2{\Gamma}}_{\cal L}$
$$
\begin{array}{lll}
&ie^\mu(\partial_\mu\Psi-\Psi a_\mu) - m(\Psi N + e^5\Psi K)=0,\\
&\ss{2}\pi(e^\mu_{\cal L}(\partial_\mu A-[A,a_\mu])-A^2)-F=0,\\
&e^\mu_{\cal L}(\partial_\mu F-[F,a_\mu])=-J(\Psi),
\end{array}\eqno(5.1)
$$
which is invariant under the gauge transformation
$$
\begin{array}{lll}
& \Psi\to\Psi^\prime=\Psi U,\\
& a_\mu\to a_\mu^\prime=U^{-1}a_\mu U+U^{-1}\partial_\mu U, \\
& F\to F^\prime=U^{-1}FU
\end{array}\eqno(5.2)
$$
where $U\in{\cal G}$, and in the right hand part of Yang-Mills equations the
value
$J(\Psi)$ is composed from the currents of Dirac's \g equation.

We hope, that the systems of equations of (5.1) type will find
applications in theoretical physics for the description of the interactions
of elementary particles.

The problem to be solved can be formulated in the following form: to find
possibly wide classes of Lie groups for which operator $J$ satisfying
conditions 1),2) can be defined. This defines systems of equations (5.1).

We may say at once what is done -- in what follows more or less complete
investigation has been done for the class of unitary Lie groups. We
have also some results for Clifford's groups which are not included in the
current paper.

\subsection{The Dirac-Yang-Mills System of \g Equations.}

Let us consider the Dirac \g equation (3.8) with \g numbers
$N,K\in\gc$ satisfying conditions
 (1.11) and (3.11) and with a gauge field
$a_\mu\in{\cal L}(t_1,\ldots,t_n)$, where generators of Lie algebra satisfy
unitary conditions
$t_k^\dagger=-t_k$ and  ADRIO condition. Let
${\cal G}\subseteq{\cal G}_{max}(N,K)$ be a gauge group of the equation (3.8) such,
that
${\cal L}(t_1,\ldots,t_n)$ is its real Lie algebra. The fact that values
$a_\mu\in{\cal L}$ in the equation
(3.8) transform under a gauge transformation by the rule
$a_\mu\to U^{-1}a_\mu U+U^{-1}\partial_\mu U,\ (U\in{\cal G})$ leads us to the
conclusion --- values $a_\mu$ must satisfy the Yang-Mills \g equations
 (4.1).

\df. A system of equations
$$
\begin{array}{lll}
&ie^\mu(\partial_\mu\Psi-\Psi a_\mu) - m(\Psi N + e^5\Psi K)=0,\\
&\ss{2}\pi(e^\mu_{\cal L}(\partial_\mu A-[A,a_\mu])-A^2)-F=0,\\
&e^\mu_{\cal L}(\partial_\mu F-[F,a_\mu])=
-\epsilon t_k\otimes\ss{1}\pi(\Psi t_k \bar\Psi i),
\end{array}\eqno(5.3)
$$
where $\Psi=\Psi(x)\in\Gamma_\C\ $, $a_\mu=a_\mu(x)=a_\mu^k
t_k\in{\cal L}\ $, $A=a_\mu e^\mu_{\cal L}\in\ss{1}\Gamma_{\cal
L}\ $, $F\in\ss{2}\Gamma_{\cal L}$, $\epsilon=1$ or
$\epsilon=-1$, is called the Dirac-Yang-Mills system of \g
equations (DYM) with a gauge group ${\cal G}$. If a gauge group
${\cal G}={\cal G}(t)$ with only one generator $t^\dagger=-t$ is
Abelian isomorphic to the group $\U(1)$ then  a system of
equations
$$
\begin{array}{lll}
&ie^\mu(\partial_\mu\Psi-\Psi t a_\mu) - m(\Psi N + e^5\Psi K)=0,\\
&\ss{2}\pi(e^\mu\partial_\mu A)-F=0,\\
&e^\mu\partial_\mu F=
-\epsilon \ss{1}\pi(\Psi t \bar\Psi i),
\end{array}\eqno(5.4)
$$
where $a_\mu\in\R$, $A=a_\mu e^\mu\in\ss{1}\Gamma$, $F\in\ss{2}\Gamma$
is called the Dirac-Maxwell system of \g equation. \par

A system of equations (5.4) is evidently invariant under the gauge
transformation
$\Psi\to\Psi\exp(\lambda t)$,
$a_\mu\to a_\mu+t\partial_\mu\lambda$, where $\lambda=\lambda(x)\in\R$.

\ti2 Theorem 1. A system of equations (5.3) is invariant under the gauge
transformation
$$
\begin{array}{lll}
& \Psi\to\Psi^\prime=\Psi U,\\
& a_\mu\to a_\mu^\prime=U^{-1}a_\mu U+U^{-1}\partial_\mu U, \\
& F\to F^\prime=U^{-1}FU
\end{array}\eqno(5.5)
$$
where $U\in{\cal G}$.\par

\proof of the gauge invariance of the first equation in (5.3) has been done in
the theorem 3, section 3. A proof of gauge invariance of the second equation
and the third equation with zero right hand part has been done in the theorem
1, section 4. So, the only thing to do for us is to prove a gauge
invariance of the right hand part
$J(\Psi)=-\epsilon t_k\otimes\ss{1}\pi(\Psi t_k \bar\Psi i)$ of the
Yang-Mills system of \g equations. This invariance
 $J(\Psi U)=U^{-1}J(\Psi)U$ follows from the chain of identities
$(\epsilon=-1)$
$$
\begin{array}{lll}
J(\Psi U)&=t_k\otimes\ss{1}\pi(\Psi Ut_k U^\dagger\bar\Psi i)=
t_k\otimes\ss{1}\pi(\Psi\{\Ad(U)\}_{lk}t_l \bar\Psi i)\cr
&=(\{\Ad(U)\}_{lk}t_k)\otimes\ss{1}\pi(\Psi t_l \bar\Psi i)=
(U^\dagger t_l U)\otimes\ss{1}\pi(\Psi t_l \bar\Psi i)=U^{-1}J(\Psi)U.
\end{array}$$
Here we have used the unitary condition $U^{-1}=U^\dagger$ and the fact,
that generators of Lie algebra
${\cal L}(t_1,\ldots,t_n)$ satisfy ADRIO condition and hence (theorem 8, section
2)
$$
Ut_k U^\dagger=\{\Ad(U)\}_{lk}t_l,\quad
U^\dagger t_l U=\{\Ad(U)\}_{lk}t_k\fin
$$

\note  Using the theorem about decomposition of the Dirac \g equation
(theorem 2, section 3), we may conclude that if in the system of equations
(5.3) \g numbers $N,K\in\dot\gc$ and the gauge group ${\cal G}\subseteq\ddot\gc$,
then in (5.3) we may take
$\Psi\equiv\ddot\Psi$, or $\Psi\equiv\dot\Psi$.
\par

\subsection{The Case $m=0$.}

There are two identities
(3.13) and (3.14) in lemma 1 section 3, from which we may hope to derive
the currents of the Dirac \g equation. In the system of \g equations (5.3)
at the right hand part of the Yang-Mills \g equations we have taken the
value
$J(\Psi)=\epsilon t_k\otimes\ss{1}\pi(\Psi t_k\bar\Psi i)$, which is composed
from the currents of the Dirac \g equation
 (3.1) (theorem 4, section 3). These currents can be derived only from the
 identity
 (3.13) and not from (3.14). Such a choice of  $J(\Psi)$ is not random,
 because in case
$m\neq0$ one can not derive any currents from the identity (3.14). The
situation is different when $m=0$. In this case each identity
(3.13) and (3.14) gives us 16 independent currents accordingly. That means,
that as a right hand part of the Yang-Mills system of \g equations (5.3)
we may take not only a value
$J(\Psi)$, but also the value
$$
\tilde J(\Psi)=-\epsilon t_k\otimes\ss{1}\pi(\Psi t_k\bar\Psi e^5),\eqno(5.6)
$$
which is composed from the currents of the Dirac \g equation that are derived
from the formula (3.14). This can be shown with the aid of identities
$$
\ss{1}\pi(\Psi t_k\bar\Psi e^5)=-g_{\mu\nu}\tilde j^\mu_k e^\nu,\eqno(5.7)
$$
$$
\ss{{\cal L}}\pi(\bar\Psi e^5 e^\mu\Psi)=\tilde j^\mu_k t_k \eqno(5.8)
$$
(compare them with the formulas (3.12) and (3.18)). So, in case $m=0$, a
right hand part of the Yang-Mills \g equations (5.3) is defined not
uniquely and this fact leads to the important consequences, which we shall
consider just now.

Let us introduce \g numbers $L=(e+ie^5)/2\ $, $R=(e-ie^5)/2$ with the
properties
$$
L+R=e,\quad L-R=ie^5,\quad L^2=L,\quad R^2=R,\quad LR=RL=0
$$
and denote $\Psi_L=L\Psi\ $, $\Psi_R=R\Psi$. Then
$$
\Psi=\Psi_L+\Psi_R,\quad \Psi_L=L\Psi_L,\quad\Psi_R=R\Psi_R,\quad
L\Psi_R=R\Psi_L=0,
$$
It is easy to calculate, that
$$
\begin{array}{lll}
\bar\Psi e^5 e^\mu\Psi+\bar\Psi i e^\mu\Psi&=
2\overline{(\Psi_L)}e^\mu\Psi_L,\cr
\bar\Psi e^5 e^\mu\Psi-\bar\Psi i e^\mu\Psi&=2\overline{(\Psi_R)}e^\mu\Psi_R.
\end{array}$$
Therefore, in case $m=0$ we get from the formulas
(3.13),(3.14)
$$
\partial_\mu(\overline{(\Psi_L)}e^\mu\Psi_L)=0,\quad
\partial_\mu(\overline{(\Psi_R)}e^\mu\Psi_R)=0.\eqno(5.9)
$$
In the same way, if we take a sum and a reminder of the right hand parts of
(3.18) and
(5.8), we get the identities
$$
\ss{{\cal L}}\pi(\overline{(\Psi_L)}ie^\mu\Psi_L)=j^{\mu+}_k t_k,\quad
\ss{{\cal L}}\pi(\overline{(\Psi_R)}ie^\mu\Psi_R)=j^{\mu-}_k t_k,\eqno(5.10)
$$
where $j^{\mu+}_k=(j^\mu_k+\tilde j^\mu_k)/2\ $,
$j^{\mu-}_k=(j^\mu_k-\tilde j^\mu_k)/2$.
It is not difficult to calculate also a sum and a reminder of the right
hand parts of
(3.12) and (5.7) and get
$$
\ss{1}\pi(\Psi_L t_k\overline{(\Psi_L)} i)=-g_{\mu\nu} j^{\mu+}_k e^\nu,\quad
\ss{1}\pi(\Psi_R t_k\overline{(\Psi_R)} i)=-g_{\mu\nu} j^{\mu-}_k e^\nu.
$$
So, in case $m=0$, we can use as a right hand part of (5.3) the value
$J(\Psi_L)=\epsilon t_k\otimes\ss{1}\pi(\Psi_L t_k\overline{(\Psi_L)} i)$,
which is composed from currents of the equation $e^\mu\partial_\mu\Psi_L=
0$,
and also we can use the value
$J(\Psi_R)$,  which is composed from the currents of the equation
$e^\mu\partial_\mu\Psi_R=0$. In the first case we can multiply the Dirac
\g equation in (5.3) from left to $R$, and using the identity
$R e^\mu=e^\mu L$, get a system of equations
$$
\begin{array}{lll}
&ie^\mu(\partial_\mu\Psi_L-\Psi_L a_\mu) =0,\\
&\ss{2}\pi(e^\mu_{\cal L}(\partial_\mu A-[A,a_\mu])-A^2)-F=0,\\
&e^\mu_{\cal L}(\partial_\mu F-[F,a_\mu])=
-\epsilon t_k\otimes\ss{1}\pi(\Psi_L t_k \overline{(\Psi_L)} i),
\end{array}\eqno(5.11)
$$
that is dependent only on
$\Psi_L$ and independent of $\Psi_R$. Similarly we can get a system of
equations for
$\Psi_R$. The systems of equations for
$\Psi_L$ and $\Psi_R$ are independent and more than this, they may have
different gauge groups (in the Standard model of electroweak interactions
such a situation takes place in the description of neutrino).

We can get one more example of a gauge invariant system of equations if, in
case $m=0$, we take a gauge group
${\cal G}$ from the real Clifford algebra $\Gamma$. Such gauge groups, in
particular, are
$$
\begin{array}{lll}
&{\cal
G}(e^1,e^2,e^3,e^{12},e^{13},e^{23},e^{012},e^{013},e^{023},e^5)
\sim\Sp(2),\\
&{\cal
G}(e^{12},e^{13},e^{23},e^{012},e^{013},e^{023})\sim\SU(2) \times\SU(2),\\
&{\cal G}(e^{12},e^{13},e^{23})\sim\SU(2),\\
&{\cal G}(e^{12})\sim\U(1).\end{array}
\eqno(5.11a)
$$
In that case we have a gauge invariant system of equations for
$\Psi$ from the real Clifford algebra $\Gamma$:
$$
\begin{array}{lll}
&e^\mu(\partial_\mu\Psi-\Psi a_\mu) =0,\\
&\ss{2}\pi(e^\mu_{\cal L}(\partial_\mu A-[A,a_\mu])-A^2)-F=0,\\
&e^\mu_{\cal L}(\partial_\mu F-[F,a_\mu])=
-\epsilon t_k\otimes\ss{1}\pi(\Psi t_k \bar\Psi e^5).
\end{array}\eqno(5.11b)
$$
In the case of last two groups from (5.11a) we may take
$\Psi\in\ddot\Gamma$, or $\Psi\in\dot\Gamma$ in (5.11b).

\subsection{A Gauge Group $\U(4)$. The Polar Gauge.}

Let us consider a system of equations (5.3) with
$\Psi N + e^5 \Psi K=\exp(\phi e^5)\Psi,\ \phi\in\R$ and with the gauge
group $\U(4)$. And let
$\Psi=\Psi(x)\in\gc,\ A=A(x)\in{\ss{1}{\Gamma}}_{\cal L},\
F=F(x)\in{\ss{2}{\Gamma}}_{\cal L}$ be a continuously
differentiable solution of this system of equations. There is a
theorem about a polar decomposition of the square matrix.

\ti2 Theorem 2 \hbox{[11]}.  Every matrix $A\in\M(n,\C)$ can be written as
$A=P U$, where $P\in\M(n,\C)$ is a hermitian nonnegatively defined matrix of
the same rank as $A$, and a matrix $U\in\M(n,\C)$ is unitary.\par

If we have a solution of the system of \g equations DYM
$\Psi=\Psi(x)\in\gc$ in representation (1.4) as a matrix from
$\M(4,\C)$, then, by the theorem 2, at each point  $x\in\R^4$ matrix
 $\Psi$ can be written in a form
$\Psi=P U$, where $P$ -- hermitian nonnegatively defined matrix from
$\M(4,\C)$, $U$ -- unitary matrix from $\M(4,\C)$.
Let us suppose, that matrix $\Psi=\Psi(x)$ can be represented in the form
$\Psi(x)=P(x) U(x)$ not only in every point $x$, but also in some domain
$\Omega\subset\R^4$ in such a way, that $P(x), U(x)$ have continuously
differentiable elements for all
$x\in\Omega$ (the possibility of the polar decomposition in domain is not
evident and need an additional investigation). In that case the solution
$\Psi=\Psi(x)$ of (5.3) in
$\Omega\subset\R^4$ defines unitary matrix
$U=U(x)$ with continuously differentiable coefficients. A matrix $U$, by
the isomorphism
$\M(4,\C)\sim\gc$ define a \g number
$U=U(x)$ from $\gc$. Let us take a \g number
$V=V(x)=U^{-1}$. A system of equations (5.3) is invariant with respect
to
the gauge group
$\U(4)$, and so, after the gauge transformation with the aid of
$V\in\U(4)$ we get such a solution of (5.3)
$$
\begin{array}{lll}\Psi^\prime&=\Psi V,\cr
a_\mu^\prime&=V^{-1}a_\mu V+V^{-1}\partial_\mu V,\cr
F^\prime&=V^{-1} F V
\end{array}$$
that in domain  $\Omega$ \g number $\Psi^\prime=\Psi^\prime(x)$ is a
hermitian nonnegatively defined \g number (in representation (1.4) it
corresponds to hermitian nonnegatively defined matrix).

\df. If $\Psi\in\gc,\ A\in{\ss{1}{\Gamma}}_{\cal L},\
F\in{\ss{2}{\Gamma}}_{\cal L}$ is such a solution of the system of \g equations
 DYM (5.3) with a gauge group
$\U(4)$, that $\Psi=\Psi(x)$ is a hermitian nonnegatively defined \g
number, then we shall call $\Psi,A,F$ a solution of
DYM in polar gauge.\par

In correspondence with the generally adopted interpretation of quantum
mechanics, particles are described by the wave functions, which are vectors
in some complex finite dimensional or infinite dimensional vector (Hilbert)
space. Observables are hermitian operators on that space. It will be
essential to suppose, that the solution $\Psi$ of (5.3) describes a wave
function of a fermion in 16 -- dimensional complex vector space. If we
consider $\Psi$ in polar gauge (in the case of gauge group $\U(4)$), then a
\g number $\Psi$ corresponds to some hermitian $4\times4$-matrix, or
hermitian
$16\!\times\!16$-matrix $\Psi\otimes{\bf1}$. That means,  $\Psi$ can be
considered not only as a wave function, but simultaneously, as some
observable value. This situation needs further consideration.

\subsection{The Gauge Group
$\U(1)\times\U(1)\times\U(1)\times\U(1)$. Spinors.}

Let
${\cal M}(4,\C)$ be an algebra of complex matrices of fourth order and
${\cal H}$ is its maximal commutative subalgebra. An algebra
${\cal H}$ is isomorphic to
the algebra of diagonal matrices with basis matrices
$s^{(1)},s^{(2)},s^{(3)},s^{(4)}$ with only nonzero element
$s^{(k)}_{kk}=1$. For this basis
$$
s^{(1)}+s^{(2)}+s^{(3)}+s^{(4)}={\bf1},\quad (s^{(k)})^2=s^{(k)},\quad
s^{(k)}s^{(l)}=0\quad\ \hbox{when}\ k\neq l. \eqno(5.12)
$$
An arbitrary matrix $B\in{\cal M}(4,\C)$ can be written in a form
$$
B=B^{(1)}+B^{(2)}+B^{(3)}+B^{(4)},\quad B^{(k)}=Bs^{(k)},\quad
B^{(k)}s^{(k)}=B^{(k)} \eqno(5.13)
$$
That means, that ${\cal M}(4,\C)$ represents as a direct sum of
its subalgebras\hfill\break
${\cal M}(4,\C)={\cal I}^{(1)}\oplus{\cal I}^{(2)}\oplus{\cal
I}^{(3)}\oplus{\cal I}^{(4)}$.  Algebras ${\cal I}^{(k)}$ are
minimal ideals of the algebra ${\cal M}(4,\C)$ because ${\cal
M}(4,\C){\cal I}^{(k)}\subseteq{\cal I}{(k)}$.  There is no
identity element in the algebra ${\cal I}^{(k)}$, but there is
``right identity element" $s^{(k)}$. Elements $B$ of the algebra
${\cal I}^{(k)}$ can be characterized by the equality $B
s^{(k)}=B$.  We shall call them k-spinors ($k=1,2,3,4$). So,
k-spinor is a  \4-matrix with all columns except the column
number k are equal to zero.

This simple consideration of matrices can be transferred word
for word to \g numbers, if we take $\gc$ instead of ${\cal
M}(4,\C)$, $e$ instead of ${\bf1}$, and
$$
\begin{array}{lll}
s^{(1)}&=(e+e^0+i e^{12}+i e^{012})/4,\\
s^{(2)}&=(e+e^0-i e^{12}-i e^{012})/4,\\
s^{(3)}&=(e-e^0+i e^{12}-i e^{012})/4,\\
s^{(4)}&=(e-e^0-i e^{12}+i e^{012})/4.
\end{array}\eqno(5.14)
$$
k-spinors are such \g numbers $B\in\gc$, that $Bs^{(k)}=B$.
For example, a general form of 1-spinors is the following:
$$
\begin{array}{lll}B=
&b_1(e+e^0+i e^{12}+i e^{012}) +
b_2(-e^{13}+i e^{23}-e^{013}+i e^{023})+\\
&b_3(-e^3+e^{03}-i e^{123}+i e^5)+
b_4(-e^1+i e^2+e^{01}-i e^{02}),
\end{array}\eqno(5.15)
$$
where $b_1,b_2,b_3,b_4\in\C$. The considered construction of spinors as
minimal ideals of the Clifford algebra is well known to specialists.

Now, let us come back to the system of equations DYM, depending on a gauge
Lie algebra ${\cal L}$ which is a subalgebra of the Lie algebra
$\u(4)\sim\u(1)\oplus\su(4)$. Note, that $\u(4)$ is isomorphic to the set
of all antihermitian \g numbers from
$\gc$. A maximal Abelian subalgebra of the Lie algebra
$\u(4)$ is isomorphic to $\u(1)\oplus\u(1)\oplus\u(1)\oplus\u(1)$
and called Cartan's subalgebra. It is evident, that the Lie algebra
$$
{\cal L}=\{2 i b_k s^{(k)}, b_k\in\R\}\eqno(5.16)
$$
is isomorphic to   $\u(1)\oplus\u(1)\oplus\u(1)\oplus\u(1)$ and hence,
${\cal L}$ is Cartan's subalgebra of the Lie algebra $\u(4)$. Generators
$t_k=2 i s^{(k)}$ of the Lie algebra (5.16) are called spinorial
generators. They are normalized by the same identity
$\tr({t_k}^2) = -4$ as vectors of antihermitian basis
(2.8) and Gell-Mann's basis (2.20).

Let us consider a Dirac-Maxwell system of \g equations with the Abelian
gauge algebra
${\cal L}=\u(1)\oplus\u(1)\oplus\u(1)\oplus\u(1)$ and with
spinorial generators $t_k=2i s^{(k)},\ k=1,2,3,4$
$$
\begin{array}{lll}
&ie^\mu(\partial_\mu\Psi-\Psi a_\mu)-m\Psi N=0,\\
&\ss{2}{\pi}(e^\mu_{\cal L}\partial_\mu A)-F=0,\\
&e^\mu_{\cal L}\partial_\mu F=
-\epsilon t_k \otimes \ss{1}{\pi}(\Psi t_k \bar\Psi i),
\end{array}\eqno(5.17)
$$
where $\Psi\in\gc$, $\ a_\mu=a_\mu^k t_k=2 i a_\mu^k s^{(k)}$,
$\ A=a_\mu e^\mu_{\cal L}$, $\ F=\sum_{\mu<\nu} f_{\mu\nu}^k t_k
e^{\mu\nu}_{\cal L}$,
$\ \epsilon=\pm1$, $\ N= q^k s^{(k)}$,
$\ q^k=\pm1$.

\ti2 Theorem 3. The system of equations (5.17) is equivalent to four systems
of equations $(l=1,2,3,4)$
$$
\begin{array}{lll}
&ie^\mu(\partial_\mu-i a_\mu^l)\psi^{(l)}- m q^l \psi^{(l)}=0,\\
&\partial_\mu a_\nu^l-\partial_\nu a_\mu^l-f_{\mu\nu}^l=0,\\
&\partial^\mu f_{\mu\nu}^l=\epsilon j_\nu^l(\psi^{(l)}),
\end{array}\eqno(5.18)
$$
where $\psi^{(l)}$ is a column number $l$ of the corresponding (in
representation (1.4)) matrix
$\Psi$,  $j_\mu^l=\bar\psi^{(l)}e_\mu\psi^{(l)}$,
$\ e_\mu=g_{\mu\nu}e^\nu$, $\ f_{\mu\nu}^l=-f_{\nu\mu}^l$
$\mu\ge\nu$.\par

\proof We multiply from right each of three equations (5.17)
by $s^{(l)}$, $l=1,2,3,4$. As a result we get four systems of equations
that are equivalent to appropriate systems
(5.18). To prove the equivalence of the appropriate right hand parts of
Maxwell's equations we have to use formulas
(3.12) and (3.18)\fin

Let us note, that wave functions $\psi^{(l)}$ from the theorem 3 can be
regarded as wave functions of four different particles with the same mass,
each of which interacts with its own gauge field. It is essential to try to
generalize our considerations for the descriptions of four particles with
different masses.

Let us consider an equation (3.1) where
$m=1,\ [N,K]=0,\ N^2+K^2=M^2$,
$M=\diag(m_1,m_2,m_3,m_4),\ m_l\geq0$.
 In that case a solution $\Psi$ of the
equation (3.1) also satisfies an equation
$$
\dsl^2\Psi+\Psi M^2=0.
$$
Multiplying from right this equation on  $s^{(l)}$ and denoting
$\psi^{(l)}=\Psi s^{(l)}$, we come to four Klein-Gordon's equations
$$
(\dsl^2+{m_l}^2)\psi^{(l)}=0,\quad l=1,2,3,4.
$$
Now we can consider a Dirac-Maxwell system of \g equations
(5.17), where
$m=1,\ N=M$. This system is reduced to four systems of equations
$$
\begin{array}{lll}
&ie^\mu(\partial_\mu-i a_\mu^l)\psi^{(l)}- m_l \psi^{(l)}=0,\\
&\partial_\mu a_\nu^l-\partial_\nu a_\mu^l-f_{\mu\nu}^l=0,\\
&\partial^\mu f_{\mu\nu}^l=\epsilon \bar\psi^{(l)}e_\mu\psi^{(l)},
\end{array}\eqno(5.19)
$$
which describe wave functions of four spin $1/2$ particles with masses
$m_l,\ l=1,2,3,4$ interacting with the gauge fields $a_\mu^{(l)}$,
each of which can be identified with electromagnetical field. If we
suggest, that
$m_1,m_2,m_3$ are masses of electron, muon and $\tau$-particle
respectively, and
$m_4$ is a mass of yet undiscovered  heavy lepton, or
$m_4=\infty$, then we get, that \g equation (5.17) in case $m=1,\ N=M$
and equations (5.19) describe four (or three) generations of charged
leptons. In the same way we can describe four (or three) generations of
quarks, but in that case we have to use Dirac's \g equation with wave
function from
${\cal M}(3,\C)\otimes\gc$.

\vskip 1cm
Let us make two final remarks.

It can be shown, that the Dirac \g equation (3.1) is invariant under
Lorentz transformation of space-time in the same manner as a standard Dirac
equation. This question will be discussed in  next publication.\par

In our approach to the gauge fields theory a maximal gauge group of the Dirac
\g equation is
$\U(4)$, which contain subgroups
$\U(1),\SU(2),\SU(3)$ of Standard Model describing interactions of
elementary particles. One of important problem of modern theoretical physics
is to develop a variant of gauge fields theory which would unify strong and
electroweak interactions of elementary particles. We see three
possibilities to use described construction for the development of GUT.
The first possibility is to develop GUT on the basis of
$\U(4)$ gauge group. There are several variants of preon models with
$\SU(4)$ gauge symmetry. The second possibility is to suppose that physical
space has a dimension $n>4$ (like in Kaluza-Klein theory) and hence, it is
possible to use the Dirac \g equation with a gauge group bigger than
$\U(4)$. The third possibility is to develop a generalized Dirac \g
equation on the basis of algebra
${\cal M}(n,C)\otimes\gc$ instead of $\gc$. Such a
generalization will have $\U(n)$ gauge symmetry.

\end{document}